\documentclass[12pt, a4paper]{article}
\pdfoutput=1

\usepackage{amsmath,amssymb}
\usepackage{comment}
\usepackage{multirow}
\usepackage[utf8]{inputenc}
\usepackage{bm}
\usepackage{accents}
\usepackage{cite}

\usepackage{tikz}
\usepackage{tikz-feynman}
\tikzfeynmanset{compat=1.0.0}

\usepackage{ifpdf}
\ifpdf        
  \usepackage{graphicx, hyperref, xcolor}     
\else     
  \usepackage[dvipdfmx]{graphicx, hyperref, xcolor}     
 \fi
\usepackage{caption}
\usepackage{subcaption}
 
\definecolor{orangered}{HTML}{FF4500}
\definecolor{crimson}{HTML}{DC143C}
\definecolor{rossoferrari}{HTML}{D9073D}
\definecolor{steelblue}{HTML}{4682B4}
\definecolor{mediumblue}{HTML}{0000CD}
\definecolor{forestgreen}{HTML}{228B22}
\hypersetup{
setpagesize=false,
bookmarksnumbered=true,%
bookmarksopen=true,%
colorlinks=true,%
linkcolor=rossoferrari,
urlcolor=steelblue,
citecolor=steelblue,
linktocpage=false,
}

\usepackage[height=23.5cm,width=17.5cm,centering]{geometry}

\newcommand{\Mp}{M_P}

\newcommand{\DAlambert}{\mathop{}\!\mathbin\Box}

\makeatletter
\@addtoreset{equation}{section}

\makeatother

\begin{document}

\begin{titlepage}

\begin{center}

\hfill DESY 19-117

\vspace{10mm}

{\LARGE \bf
Dynamical Emergence of Scalaron \\ \vspace{6.5mm}
in Higgs Inflation
}

\vspace{10mm}

{\large
Yohei Ema
}

\vspace{5mm}

{\em DESY, Notkestra{\ss}e 85, D-22607 Hamburg, Germany}\\[.3em]

\vspace{5mm}

\end{center}

\begin{abstract}
We point out that a light scalaron dynamically emerges
if scalar fields have a sizable non-minimal coupling to the Ricci scalar
as in the Higgs inflation model.
We support this claim in two ways. 
One is based on the renormalization group equation;
the non-minimal coupling inevitably induces a Ricci scalar quadratic term
due to the renormalization group running.
The other is based on scattering amplitudes;
a scalar four-point amplitude
develops a pole after summing over a certain class of diagrams,
which we identify as the scalaron.
Our result implies that the Higgs inflation is actually a two-field inflationary model.
Another implication is that 
the Higgs inflation does not suffer from the unitarity issue
since the scalaron pushes up the cut-off scale to the Planck scale.
\end{abstract}

\end{titlepage}

\renewcommand{\thepage}{\arabic{page}}
\setcounter{page}{1}

\tableofcontents
\pagebreak
\renewcommand{\thepage}{\arabic{page}}
\renewcommand{\thefootnote}{$\natural$\arabic{footnote}}
\setcounter{footnote}{0}

\section{Introduction}
\label{sec:intro}

Inflation plays an essential role in the modern cosmology.
It not only solves the initial condition problems such as the flatness and horizon problems,
but also provides seeds for the large scale structure
and the anisotropy of the cosmic microwave background (CMB) in the present universe.

Among many candidates, 
the Higgs inflation model~\cite{Futamase:1987ua,CervantesCota:1995tz,Bezrukov:2007ep} 
is probably one of the most popular inflationary models because 
of its minimality and consistency with the CMB observation~\cite{Akrami:2018odb}.
It identifies the (radial component of the) standard model Higgs doublet as the inflaton.
The original Higgs quartic potential needs to be modified at the large field value region
to be consistent with the CMB,
and hence the following interaction
\begin{align}
	\mathcal{L}_{\xi} = \xi \left\vert H \right\rvert^2 R,
\end{align}
is introduced, where $\xi$ is a non-minimal coupling between 
the Higgs doublet $H$ and the Ricci scalar $R$.
The CMB normalization requires $\xi$ and the Higgs quartic coupling $\lambda$
to satisfy $\xi^2 \simeq 2\times 10^9\,\lambda$, 
indicating that $\xi \sim \mathcal{O}(10^4)$ 
unless $\lambda$ is tiny.
Such a large $\xi$ triggers a lot of discussions,
most importantly concerning the unitarity issue of 
the Higgs inflation~\cite{Burgess:2009ea,Barbon:2009ya,
Burgess:2010zq,Hertzberg:2010dc,Kehagias:2013mya}.
In the Higgs inflation, the tree-level unitarity is violated at the energy scale of $\mathcal{O}(\Mp/\xi)$
around the vacuum (where the Higgs field value vanishes).
It does not necessarily spoil the inflationary prediction since the scale of the violation
depends on the Higgs field value and is larger during inflation~\cite{Bezrukov:2010jz}.
Nevertheless, several issues, such as possible UV completion 
of the Higgs inflation~\cite{Lerner:2010mq,Giudice:2010ka,
Hertzberg:2011rc,Barbon:2015fla}
or violent phenomena caused by the large non-minimal coupling 
after inflation~\cite{Ema:2016dny,DeCross:2015uza,Sfakianakis:2018lzf}, 
have been discussed so far.

In this paper, we study consequences of the non-minimal coupling one step further,
assuming that $\xi \gg 1$.
In particular, we see that a scalaron degree of freedom (i.e. the spin-0 part of the metric)
becomes dynamical once we properly take into account quantum effects.
Our argument is based on two distinct (yet closely related) observations.
First, we study the renormalization group (RG) equation of this model.
It is well-known that, once we treat the gravity as an effective field theory, 
we have to introduce the Ricci scalar quadratic term,
\begin{align}
	\mathcal{L}_{\alpha_1} = \alpha_1 R^2,
\end{align}
as a counter term to renormalize divergences 
at the one-loop level~\cite{tHooft:1974toh,Donoghue:1995cz}.
It results in the RG running of $\alpha_1$.
The scalar loop contribution to the beta function is given by
\begin{align}
	\beta_{\alpha_1} &\equiv \frac{d\alpha_1}{d\ln \mu} = 
	-\frac{N_s}{1152\pi^2}\left(1+6\xi\right)^2,
\end{align}
where $N_s$ is the number of real scalar fields ($N_s = 4$ in the Higgs inflation).
It means that we cannot keep $\left\lvert\alpha_1\right\rvert \lesssim \mathcal{O}(1)$ 
for all energy scales for $\xi \gg 1$.
Since $\alpha_1$ is inversely proportional to the scalaron mass squared,
a light scalaron inevitably shows up in the theory.
This line of argument is also emphasized
in Refs.~\cite{Salvio:2015kka,Calmet:2016fsr,Ema:2017rqn,Ghilencea:2018rqg}.
Second, and more interestingly, 
we see that the scalaron dynamically emerges as an intermediate state
in 2-to-2 scattering amplitudes.\footnote{
	In this paper, we use the word ``dynamical" to indicate that
	the scalaron emerges after resumming over a certain class of Feynman diagrams
	(see Eqs.~\eqref{eq:sum} and~\eqref{eq:sum_Einstein}).
}
If the above argument based on the RG equation is true,
the unitarity issue of the Higgs inflation has to be gone after including quantum effects.
This is because, once the scalaron exists, 
the cut-off scale of the theory is pushed up to the Planck scale~\cite{Ema:2017rqn,Gorbunov:2018llf}.
Such a mechanism of healing the unitarity
is indeed discussed in Refs.~\cite{Aydemir:2012nz,Calmet:2013hia}, 
called the self-healing mechanism.
We show that the self-healed scattering amplitude
is equivalent to an amplitude with the scalaron propagating as an intermediate state.
It strongly supports that the scalaron dynamically emerges
once we take into account quantum effects properly.

This paper is organized as follows. 
In Sec.~\ref{sec:RGE}, we argue that the scalaron inevitably shows up
in the theory based on the RG equation. 
It also provides some notations used in this paper.
In Sec.~\ref{sec:self_healing}, 
we study a 2-to-2 scattering amplitude between the scalar fields.
We first review the unitarity issue of the Higgs inflation based on the scattering amplitude,
and the self-healing mechanism.
We then move to our main point, identifying the self-healing mechanism as
the dynamical emergence of the scalaron.
Sec.~\ref{sec:equivalence} shows the equivalence of our results between the Jordan and Einstein frames.
We consider only the case $N_s \geq 2$ 
in Secs.~\ref{sec:RGE}--\ref{sec:equivalence}.
The case $N_s = 1$ is tricky and is separately discussed in Sec.~\ref{sec:single}.
Note that the subtlety discussed there is irrelevant for the Higgs inflation
that has $N_s = 4$.
Finally Sec.~\ref{sec:summary} is devoted to summary and discussions.
The appendix is composed of two parts. In App.~\ref{app:details},
we provide some computational details
for the sake of clarity and completeness.
We review the Higgs scalaron inflationary model, 
i.e. an inflationary model with the Higgs and the scalaron, in App.~\ref{app:HiggsR2}.

\section{Renormalization group equation and scalaron}
\label{sec:RGE}

In this paper, we consider the following action:
\begin{align}
	S = \int d^4x \sqrt{-g}\left[
	\frac{\Mp^2}{2}R
	+ \frac{1}{2}g^{\mu\nu}\partial_\mu \phi_i \partial_\nu \phi_i
	+ \frac{\xi}{2}R\phi_i^2
	- \frac{\lambda \phi_i^4}{4}
	\right]
	+ S_\mathrm{c.t.},
	\label{eq:action}
\end{align}
where $\Mp$ is the (reduced) Planck scale, 
$\phi_i$ are the real scalar fields, $i = 1, 2, ..., N_s$, 
$g_{\mu\nu}$ is the metric with $g$ its determinant, $R$ is the Ricci scalar,
and $\xi$ and $\lambda$ are the coupling constants. 
We ignore the electroweak scale throughout this paper.
It is justified as the energy scale of our interest is significantly higher than the electroweak scale.
We focus on the case $N_s \geq 2$ in this section, 
Secs.~\ref{sec:self_healing} and~\ref{sec:equivalence}.
The Higgs inflation corresponds to $N_s = 4$.\footnote{
	As far as the beta functions or scattering amplitudes
	around the vacuum at high energy scale are concerned,
	it is appropriate to treat the Higgs doublet 
	as four real scalar fields.
}
The case $N_s = 1$ is tricky and is discussed separately in Sec.~\ref{sec:single}. 
The last term $S_\mathrm{c.t.}$ is the counter term required to renormalize
this theory. It plays an essential role in the following.

The inflationary prediction of this model (without the counter term)
is studied in detail in literature, especially in the case that $\phi_i$ are 
(the components of) the standard model Higgs doublet.
It reproduces the normalization of the CMB anisotropy when
\begin{align}
	\xi^2 \simeq 2\times 10^9\,\lambda.
\end{align}
It requires that $\xi \gg 1$ unless $\lambda$ is tiny,
and hence we assume that $\xi \gg 1$ throughout this paper.\footnote{
	The case with tiny $\lambda$ is called 
	the critical Higgs inflation model~\cite{Hamada:2014iga,Bezrukov:2014bra}.
	Our discussion in the following does not apply to this case
	as long as $\xi \lesssim \mathcal{O}(1)$.
}
The spectral index is consistent with the current CMB observation~\cite{Akrami:2018odb},
and the tensor-to-scalar ratio is within the reach of the future observations~\cite{Abazajian:2016yjj}.
Because of these features and its minimality, 
the Higgs inflation is one among the most popular inflationary models so far.

The inflationary prediction is usually studied at the classical level.\footnote{
	See Refs.~\cite{Barvinsky:2008ia,DeSimone:2008ei,
	Bezrukov:2008ej,Bezrukov:2009db,Barvinsky:2009fy,
	Barvinsky:2009ii,Allison:2013uaa,
	George:2015nza,Fumagalli:2016lls,Markkanen:2017tun} 
	and references therein for quantum effects 
	on the Higgs potential and the non-minimal coupling.
	They are also studied based on the exact RG group approach 
	in Ref.~\cite{Saltas:2015vsc}.
}
Once we consider quantum effects, however,  we have to take into account counter terms.
Since this theory contains the gravity,
the following counter terms (among others) 
are required to cancel divergences at the one-loop level~\cite{tHooft:1974toh}:
\begin{align}
	S_{\mathrm{c.t.}} = \int d^4x \sqrt{-g}\,\left[ \alpha_1 R^2 + \alpha_2\left(R_{\mu\nu}R^{\mu\nu} 
	- \frac{1}{3}R^2\right)\right].
\end{align}
As a result, the coefficients $\alpha_1$ and $\alpha_2$ run according to the energy scale
of the system. The (scalar loop contributions to the) beta functions are given by\footnote{
	Eq.~\eqref{eq:beta_alpha1} 
	agrees with Refs.~\cite{Fradkin:1981iu,Salvio:2014soa,Salvio:2015kka,
	Markkanen:2018bfx,Ghilencea:2018rqg},
	while Refs.~\cite{Codello:2015mba,Calmet:2016fsr}
	have an opposite sign.
	Note that the sign convention of the Ricci tensor 
	does not affect that of $\alpha_1$
	since the latter depends quadratically on the former.
	\label{ft:sign_beta}
}
\begin{align}
	\beta_{\alpha_1} &\equiv \frac{d\alpha_1}{d\ln \mu} = 
	-\frac{N_s}{1152\pi^2}\left(1+6\xi\right)^2, 
	\label{eq:beta_alpha1} \\
	\beta_{\alpha_2} &\equiv \frac{d\alpha_2}{d\ln \mu} =
	-\frac{N_s}{960\pi^2}.
\end{align}
See App.~\ref{app:details} for the derivation.
Two important properties are read off.
First, the runnings of $\alpha_1$ and $\alpha_2$ are additive, not multiplicative.
That is, the operators associated with $\alpha_1$ and $\alpha_2$ inevitably emerge
at other energy scales
even if they are absent at one specific energy scale.
Second, 
the size of $\alpha_1$ is generically of $\mathcal{O}(\xi^2)$
(or larger) that is significantly larger than unity since we assume $\xi \gg 1$.
It is well-known that the $R^2$ operator introduces 
an additional scalar degree of freedom~\cite{Starobinsky:1980te,Barrow:1983rx,Whitt:1984pd,Barrow:1988xh}, 
which we call ``scalaron" in this paper.
Its mass is given by
\begin{align}
	m_s^2 = \frac{\Mp^2}{12\alpha_1},
\end{align}
which is lighter as $\alpha_1$ is larger.
Thus, the beta function indicates that the Higgs inflation model with the large non-minimal coupling $\xi$
necessarily contains the scalaron as a light degree of freedom.\footnote{
	Here the word ``light" means that it is significantly lighter than the Planck scale.
} This line of argument is also emphasized 
in Refs.~\cite{Salvio:2015kka,Calmet:2016fsr,Ema:2017rqn,Ghilencea:2018rqg}.

We emphasize that we cannot simply ignore the scalaron.
In order to make the theory well-defined at the quantum level,
we have to introduce the $\alpha_1$-operator and hence the scalaron
from the beginning.
We can make the scalaron heavy at some specific energy scale by taking $\alpha_1$ small there,
but it soon becomes as light as $\mathcal{O}(\Mp/\xi)$ once one goes below that scale.\footnote{
	See Sec.~\ref{sec:summary} for more details on
	the RG running of the scalaron mass.
}
In other words, one cannot keep the scalaron heavy for all energy scales.
It is in contrast to the case of the $\alpha_2$-operator.
It is known that the $\alpha_2$-operator introduces a spin-2 ghost degree of freedom
whose mass squared is~\cite{Julve:1978xn,Salvio:2018crh}
\begin{align}
	m_2^2 = -\frac{\Mp^2}{2\alpha_2}.
	\label{eq:mass_spin2}
\end{align}
It is kept to be of $\mathcal{O}(\Mp^2)$ for all scales 
once we take $\vert\alpha_2\vert \lesssim \mathcal{O}(1)$ at some scale,
since $\beta_{\alpha_2}$ does not depend on $\xi$.
Then we might forget about it by, e.g., believing that some UV completion of the gravity
cures the pathology associated with the ghost.
The same argument does not apply to the scalaron.
Since the scalaron mass is in general significantly smaller than the Planck scale for $\xi \gg 1$,
it has to be distinguishable from the effects of the UV completion of the gravity.

Although this argument might be already convincing, 
we will support it from a different point of view in the following,
since the statement that the scalaron inevitably exists 
has a huge impact on 
both the theoretical and phenomenological sides of the Higgs inflation.
To be specific, in Sec.~\ref{sec:self_healing}, we consider a 2-to-2 scattering amplitude 
$\phi_i \phi_i \rightarrow \phi_j \phi_j$ with $i\neq j$.
We will see that the scalaron dynamically emerges
as an intermediate state in the amplitude
even if we do not include it at the beginning.
It also reveals that 
the dynamical emergence of the scalaron
is closely related to the unitarity structure of the Higgs inflation.

\section{Self-healing mechanism and scalaron}
\label{sec:self_healing}
In this section, we consider the 2-to-2 scattering amplitude $\phi_i \phi_i \rightarrow \phi_j \phi_j$
with $i \neq j$ around the vacuum $\phi_i = 0$,
and show that a scalaron degree of freedom dynamically emerges 
as an intermediate state in this process.
This phenomena is closely related to the self-healing mechanism discussed in 
Refs.~\cite{Aydemir:2012nz,Calmet:2013hia}.
Since the self-healing mechanism is studied in the context of the unitarity issue
in the Higgs inflation, we first review the latter in Sec.~\ref{subsec:unitarity}.
In Sec.~\ref{subsec:self_healing}, we review the self-healing mechanism.
The self-healing mechanism in practice sums over a certain class of Feynman diagrams
(see Eqs.~\eqref{eq:sum} and~\eqref{eq:sum_Einstein}).
As a result, the scattering amplitude develops a pole structure.
We see in Sec.~\ref{subsec:pole} that it
can be identified with the scalaron propagating as an intermediate state.

\subsection{Unitarity of Higgs inflation}
\label{subsec:unitarity}

As we have discussed in Sec.~\ref{sec:RGE}, the Higgs inflation model requires $\xi \gg 1$.
It triggers huge discussions, especially concerning the unitarity of the Higgs inflation model.
Let us consider the tree-level 2-to-2 scattering amplitude $\phi_i \phi_i \rightarrow \phi_j \phi_j$ with $i \neq j$
in the Higgs inflation. It is given as (see App.~\ref{app:details} for the derivation)
\begin{align}
	A_\mathrm{tree}^{(ii \rightarrow jj)} 
	= \frac{1}{\Mp^2s}\left[\frac{\left(1+6\xi\right)^2}{6} s^2 - \left(\frac{s^2}{6}-tu\right)\right],
	\label{eq:amp_tree}
\end{align}
around $\phi_i = 0$, where $s,t$ and $u$ are the standard Mandelstam variables.
Here and henceforth we ignore the contribution from the Higgs potential 
since it is irrelevant for our discussion.
It indicates that the cut-off of the theory is $\mathcal{O}(\Mp/\xi)$,
significantly lower than the Planck scale~\cite{Burgess:2009ea,Barbon:2009ya,
Burgess:2010zq,Hertzberg:2010dc,Kehagias:2013mya}.
It does not necessarily spoil the inflationary prediction of this model
since the cut-off scale depends on the Higgs field value,
and is higher than the typical energy scale during inflation~\cite{Bezrukov:2010jz}.
It still seems to require that the Higgs inflation should be UV completed below the Planck scale.
More interestingly, the unitarity could be violated 
during preheating~\cite{Ema:2016dny,DeCross:2015uza,Sfakianakis:2018lzf},
resulting in the necessity of UV completion to describe the reheating dynamics.
These are the (very) rough sketch of the unitarity issue on the Higgs inflation
in literature so far.

In this paper, our viewpoint on the unitarity of the Higgs inflation is totally different.
In Sec.~\ref{sec:RGE}, we argue that the scalaron inevitably shows up
in the Higgs inflation model due to the quantum effects.
If it is true, the above discussion on the unitarity has to be just an illusion, 
and the unitarity of the Higgs inflation has to be remedied below the Planck scale 
once quantum effects are properly taken into account.
This is because the cut-off scale of the Higgs inflation
is pushed up to the Planck scale once the scalaron is included,
as first pointed out in Ref.~\cite{Ema:2017rqn} 
and further studied in Ref.~\cite{Gorbunov:2018llf} (see also App.~\ref{subsec:Einstein_HiggsR2}).
Indeed, such a healing mechanism of the unitarity, or the self-healing mechanism, 
is discussed in Refs.~\cite{Aydemir:2012nz,Calmet:2013hia}.
In particular, Ref.~\cite{Calmet:2013hia} discusses the self-healing mechanism of the Higgs inflation,
concluding that it has no unitarity issue once quantum effects are included.
In the following, we shed a new light on the self-healing mechanism;
we point out that the self-healing mechanism in the Higgs inflation
can be understood as dynamical emergence of the scalaron.
In Sec.~\ref{subsec:self_healing}, we review the self-healing mechanism
 following Refs.~\cite{Aydemir:2012nz,Calmet:2013hia}.
We then move on to our main point in Sec.~\ref{subsec:pole}, 
identifying the self-healed scattering amplitude as the emergence of the scalaron.

\subsection{Self-healing mechanism}
\label{subsec:self_healing}
In this subsection, we review the self-healing mechanism discussed 
in Refs.~\cite{Aydemir:2012nz,Calmet:2013hia} (see also Ref.~\cite{Han:2004wt}).
Since the self-healing mechanism is best described by the large $N_s$ limit
with $N_s \Mp^{-2}$ fixed, we take this limit in this and the next subsections.
We focus on the 2-to-2 scattering amplitude 
$\phi_i \phi_i \rightarrow \phi_j \phi_j$ with $i \neq j$.
The tree-level amplitude is given by Eq.~\eqref{eq:amp_tree}, i.e.,
\begin{align}
	A_\mathrm{tree}^{(ii \rightarrow jj)} = \frac{1}{\Mp^2s}\left[\frac{\left(1+6\xi\right)^2}{6} s^2 - \left(\frac{s^2}{6}-tu\right)\right].
\end{align}
As we have discussed earlier, it violates the tree-level unitarity for $\sqrt{s} \gtrsim \Mp/\xi$.
This scale is usually interpreted as the scale at which the theory is UV completed.
Ref.~\cite{Aydemir:2012nz} points out, however, 
that the scale of the tree-level unitarity violation
and the onset of new physics can have a different dependence on, e.g.,
the number of flavors ($N_s$ in our case). 
In other words, the scale of the tree-level unitarity violation can be
parametrically smaller than the onset of new physics, depending on the number of flavors.
In such a case, the tree-level unitarity violation has to be remedied within the low energy theory itself,
without help of new physics.\footnote{
	Here we follow the terminology of Ref.~\cite{Aydemir:2012nz}.
	The usage of ``new physics" might be somewhat confusing,
	since the self-healing mechanism often accompanies a new dynamical degree of freedom
	(the scalaron in our case).
	Nevertheless it has to be distinguished from introducing a new particle simply by hand,
	since the new degree of freedom results from resummation of diagrams 
	that are within the low energy theory itself.
}
Ref.~\cite{Aydemir:2012nz} 
studies QCD and the electroweak gauge theory as concrete examples,
but it suggests that the Higgs inflation could also be within this class of theories,
as we now review.

In our model, the scalar loop to the graviton vacuum polarization
gives the leading order correction 
to the scattering amplitude (in the sense of the large $N_s$ limit).
It is given at the one-loop level by (see App.~\ref{app:details} for the derivation)
\begin{align}
	A_\mathrm{1\mathchar`- loop}^{(ii\rightarrow jj)} &= 
	\begin{tikzpicture}[baseline=(c)]
	\begin{feynman}[inline]
		\vertex (c);
		\vertex [above left = of c] (a);
		\vertex [below left = of c] (b);
		\vertex [right = of c] (d);
		\vertex [right = of d] (e);
		\vertex [right = of e] (f);
		\vertex [above right = of f] (g);
		\vertex [below right = of f] (h);
		\diagram*{
		(a) -- [scalar] (c) -- [scalar] (b)
		(c) -- [photon] (d)
		(e) -- [scalar, half left] (d) -- [scalar, half left] (e)
		(e) -- [photon] (f)
		(g) -- [scalar] (f) -- [scalar] (h) 
		};
	\end{feynman}
	\end{tikzpicture}
	\nonumber \\
	=& -\frac{N_s}{960\pi^2\Mp^4}\left[\frac{5}{6}
	\left(1+6\xi\right)^4s^2\left(\ln\left(\frac{s}{\mu_1^2}\right) - i \pi \right) 
	+ \left(\frac{s^2}{6}-tu\right)\left(\ln\left(\frac{s}{\mu_2^2}\right) - i \pi \right)\right],
\end{align}
where the dashed (wavy) line denotes the scalar (the graviton).
We have renormalized the divergences by the $\alpha_1$- and $\alpha_2$-operators, i.e.,
\begin{align}
	S_{\mathrm{c.t.}} = \int d^4x \sqrt{-g}\,\left[ \alpha_1 R^2 + \alpha_2\left(R_{\mu\nu}R^{\mu\nu} 
	- \frac{1}{3}R^2\right)\right],
\end{align}
and the energy scales $\mu_1$ and $\mu_2$ 
correspond to the parameter choices of our theory.
Note that we have to specify the values of $\alpha_1$ and $\alpha_2$ at some energy scale
after the renormalization.
It is equivalent to specifying the energy scales at which $\alpha_1$ and $\alpha_2$ vanish,
which we denote as $\mu_1$ and $\mu_2$, respectively.
The one-loop contribution is larger than the tree-level one 
for $\sqrt{s} \gtrsim \Mp/\sqrt{N_s}\,\xi$,
and hence we may define a dressed amplitude
by summing over the scalar loop diagrams as
\begin{align}
	A_{\mathrm{dressed}}^{(ii\rightarrow jj)} &\equiv
	\begin{tikzpicture}[baseline=(c)]
	\begin{feynman}[inline]
		\vertex (c);
		\vertex [above left = of c] (a);
		\vertex [below left = of c] (b);
		\vertex [right = of c] (d);
		\vertex [above right = of d] (e);
		\vertex [below right = of d] (f);
		\diagram*{
		(a) -- [scalar] (c)
		(b) -- [scalar] (c)
		(c) -- [photon] (d)
		(d) -- [scalar] (e)
		(d) -- [scalar] (f)
		};
	\end{feynman}
	\end{tikzpicture}
	+
	\begin{tikzpicture}[baseline=(c)]
	\begin{feynman}[inline]
		\vertex (c);
		\vertex [above left = of c] (a);
		\vertex [below left = of c] (b);
		\vertex [right = of c] (d);
		\vertex [right = of d] (e);
		\vertex [right = of e] (f);
		\vertex [above right = of f] (g);
		\vertex [below right = of f] (h);
		\diagram*{
		(a) -- [scalar] (c) -- [scalar] (b)
		(c) -- [photon] (d)
		(e) -- [scalar, half left] (d) -- [scalar, half left] (e)
		(e) -- [photon] (f)
		(g) -- [scalar] (f) -- [scalar] (h) 
		};
	\end{feynman}
	\end{tikzpicture}
	+
	\begin{tikzpicture}[baseline=(c)]
	\begin{feynman}[inline]
		\vertex (c);
		\vertex [above left = of c] (a);
		\vertex [below left = of c] (b);
		\vertex [right = of c] (d);
		\vertex [right = of d] (e);
		\vertex [right = of e] (f);
		\vertex [right = of f] (g);
		\vertex [right = of g] (h);
		\vertex [above right = of h] (i);
		\vertex [below right = of h] (j);
		\diagram*{
		(a) -- [scalar] (c) -- [scalar] (b)
		(c) -- [photon] (d)
		(e) -- [scalar, half left] (d) -- [scalar, half left] (e)
		(e) -- [photon] (f)
		(g) -- [scalar, half left] (f) -- [scalar, half left] (g)
		(g) -- [photon] (h)
		(i) -- [scalar] (h) -- [scalar] (j) 
		};
	\end{feynman}
	\end{tikzpicture}
	\nonumber \\
	&+ 
	\begin{tikzpicture}[baseline=(c)]
	\begin{feynman}[inline]
		\vertex (c);
		\vertex [above left = of c] (a);
		\vertex [below left = of c] (b);
		\vertex [right = of c] (d);
		\vertex [right = of d] (e);
		\vertex [right = of e] (f);
		\vertex [right = of f] (g);
		\vertex [right = of g] (h);
		\vertex [right = of h] (i);
		\vertex [right = of i] (j);
		\vertex [above right = of j] (k);
		\vertex [below right = of j] (l);
		\diagram*{
		(a) -- [scalar] (c) -- [scalar] (b)
		(c) -- [photon] (d)
		(e) -- [scalar, half left] (d) -- [scalar, half left] (e)
		(e) -- [photon] (f)
		(g) -- [scalar, half left] (f) -- [scalar, half left] (g)
		(g) -- [photon] (h)
		(i) -- [scalar, half left] (h) -- [scalar, half left] (i)
		(i) -- [photon] (j)
		(k) -- [scalar] (j) -- [scalar] (l) 
		};
	\end{feynman}
	\end{tikzpicture}
	+
	\cdots.
	\label{eq:sum}
\end{align}
Note that it is
the leading order contribution of the large $N_s$ expansion,
since the loops involving the graviton and ghost (from the gauge fixing) 
are suppressed by $N_s$
compared to the scalar loops.

Now we study the unitarity structure of this dressed amplitude.
We consider an elastic scattering amplitude between flavor-singlet states,
\begin{align}
	\left\vert \bm{1} \right\rangle \equiv \frac{1}{\sqrt{N_s}}\sum_{i = 1}^{N_s} \left \lvert \phi_i \phi_i \right\rangle,
	\label{eq:singlet}
\end{align}
where $\left\lvert \phi_i \phi_i\right\rangle$ denotes the state with two $\phi_i$ particles.
It is given at the leading order in $1/N_s$ by
\begin{align}
	A^{(\bm{1}\rightarrow \bm{1})} = N_s A^{(ii \rightarrow jj)}_\mathrm{dressed},
\end{align}
as the diagonal parts are suppressed by $1/N_s$.
The factor $N_s$ originates from the combinatory factor $N_s^2$ divided by the normalization
of the state (the prefactor in Eq.~\eqref{eq:singlet}).
We now perform the partial wave expansion,\footnote{
	We extract the factor $32\pi$ instead of $16\pi$
	since the final state particles are identical bosons.
	The unitarity requires
	$\mathrm{Im}\left[a^{(l)}\right] = \vert a^{(l)}\vert^2$
	with this convention in our case,
	since the cross section is divided 
	by the combinatory factor of the identical final state particles.
}
\begin{align}
	A^{(\bm{1}\rightarrow \bm{1})} = 32\pi \sum_{l}\left(2l+1\right) a^{(l)}(s) P_l \left(\cos\theta\right),
\end{align}
where $l$ is the angular momentum,
$P_l$ is the Legendre polynomial of degree $l$ with $P_l(1) = 1$, 
and $\theta$ is the scattering angle.
The Mandelstam variables $t$ and $u$ are given in terms of $s$ and $\theta$ as
\begin{align}
	t = -\frac{s}{2}\left(1 - \cos \theta\right),~~~
	u = -\frac{s}{2}\left(1 + \cos \theta \right).
\end{align}
Since the different spin parts do not mix with each other in the summation~\eqref{eq:sum},
we obtain the $s$-wave part of the partial wave amplitude as
\begin{align}
	a^{(0)}
	= \frac{a^{(0)}_\mathrm{tree}}{1 - a^{(0)}_\mathrm{1\mathchar`-loop}/a^{(0)}_\mathrm{tree}},
	\label{eq:amp_self_healed}
\end{align}
where the $s$-wave parts of the tree-level and one-loop amplitudes are respectively given as
\begin{align}
	a^{(0)}_\mathrm{tree} &= \frac{N_s\left(1+6\xi\right)^2}{192 \pi \Mp^2} s, \\
	a^{(0)}_\mathrm{1\mathchar`-loop}
	&= -\frac{N_s^2\left(1+6\xi\right)^4}{36864\pi^3 \Mp^4} s^2 
	\left[\ln \left(\frac{s}{\mu_1^2}\right) - i\pi \right].
\end{align}
We can check that it exactly satisfies the elastic unitarity,
\begin{align}
	\mathrm{Im}\left[a^{(0)}\right] = \left\lvert a^{(0)}\right\rvert^2.
\end{align}
In this sense, the unitarity is maintained at the leading order in $1/N_s$.
This is the self-healing mechanism discussed in Refs.~\cite{Aydemir:2012nz,Calmet:2013hia}.
In the following, 
we call the dressed amplitude as the self-healed amplitude.

\subsection{Self-healed amplitude as scalaron emergence}
\label{subsec:pole}

In the previous subsection, we see that the ($s$-wave part of the) self-healed amplitude
satisfies the elastic unitarity.
Now we study its structure in more detail. The $s$-wave part is given by
\begin{align}
	a^{(0)} = -\frac{N_s\left(1+6\xi\right)^2 s}{2304\pi \alpha_1}
	\left[s\left(1-\frac{i\pi}{\ln\left(s/\mu_1^2\right)}\right) - \frac{\Mp^2}{12\alpha_1}\right]^{-1},
	\label{eq:self_heal_amp}
\end{align}
where we have defined
\begin{align}
	\alpha_1 \equiv -\frac{N_s \left(1+6\xi\right)^2}{2304\pi^2}\ln \left(\frac{s}{\mu_1^2}\right).
	\label{eq:alpha1_self_heal}
\end{align}
It is indeed understood as the coupling $\alpha_1$ at the energy scale $\sqrt{s}$,
since it is consistent with the beta function $\beta_{\alpha_1}$ with
$\mu_1$ being the energy scale at which $\alpha_1$ vanishes.
We ignore the imaginary part of the amplitude for now, whose meaning will be clarified soon.
The amplitude is then written as
\begin{align}
	a^{(0)} = -\frac{N_s\left(1+6\xi\right)^2}{2304\pi \alpha_1}
	\left[\frac{m_s^2}{s-m_s^2} +1 \right],
\end{align}
where we have defined
\begin{align}
	m_s^2 \equiv \frac{\Mp^2}{12\alpha_1}.
\end{align}

We now see that it is equivalent to the tree-level scattering amplitude 
with the scalaron as a dynamical degree of freedom.
We consider the following action:
\begin{align}
	S = \int d^4 x \sqrt{-{g}} 
	\left[
	\frac{\Mp^2}{2}{R} + \bar{\alpha}_1 {R}^2
	+ \frac{\xi}{2} {R} {\phi}_i^2
	+ \frac{1}{2}{g}^{\mu\nu}\partial_\mu {\phi}_i \partial_\nu {\phi}_i
	- \frac{\lambda}{4}\phi_i^4
	\right].
\end{align}
By introducing an auxiliary field and performing the Weyl transformation,
it is equivalent to (see App.~\ref{subsec:Einstein_HiggsR2} for details)
\begin{align}
	S = \int d^4 x \sqrt{-g}
	&\left[
	\frac{\Mp^2}{2} R + \frac{1}{2} \left(\partial s\right)^2
	+ \frac{1}{2}\left(\partial \phi_i + \frac{1}{\sqrt{6}}\frac{\phi_i}{\Mp}\partial s\right)^2
	\right. \nonumber \\
	&\left.
	-\frac{\Mp^4}{16 \bar{\alpha}_1} 
	\left(1 - \exp\left(-\sqrt{\frac{2}{3}}\frac{s}{\Mp}\right) - \frac{\xi \phi_i^2}{\Mp^2}\right)^2
	- \frac{\lambda}{4} \phi_i^4
	\right],
	\label{eq:Einstein_HiggsR2}
\end{align}
where $s$ is the scalaron degree of freedom 
that shows up due to the $\bar{\alpha}_1$-term as advertised.\footnote{
	It should not be confused with the Mandelstam variable.
} We compute the 2-to-2 scattering amplitude $\phi_i \phi_i \rightarrow \phi_j \phi_j$ with $i \neq j$ 
around the vacuum $\phi_i = s = 0$ in this model
(see App.~\ref{subsec:Feynman_HiggsR2} for the Feynman rules).
The tree-level amplitude is diagrammatically given by
\begin{align}
	i\bar{A}_{\mathrm{tree}}^{(ii\rightarrow jj)}
	&= 
	\begin{tikzpicture} [baseline=(c)]
	\begin{feynman} [inline]
		\vertex (c);
		\vertex [above left = of c] (a);
		\vertex [below left = of c] (b);
		\vertex [right = of c] (d);
		\vertex [above right = of d] (e);
		\vertex [below right = of d] (f);
		\diagram*{
		(a) -- [scalar] (c) -- [scalar] (b)
		(c) -- (d)
		(e) -- [scalar] (d) -- [scalar] (f)
		};
	\end{feynman}
	\end{tikzpicture}
	+
	\begin{tikzpicture} [baseline=(c)]
	\begin{feynman} [inline]
		\vertex (c);
		\vertex [above left = of c] (a);
		\vertex [below left = of c] (b);
		\vertex [right = of c] (d);
		\vertex [above right = of d] (e);
		\vertex [below right = of d] (f);
		\diagram*{
		(a) -- [scalar] (c) -- [scalar] (b)
		(c) -- [photon] (d)
		(e) -- [scalar] (d) -- [scalar] (f)
		};
	\end{feynman}
	\end{tikzpicture}
	+
	\begin{tikzpicture} [baseline=(c)]
	\begin{feynman} [inline]
		\vertex (c);
		\vertex [above left = of c] (a);
		\vertex [below left = of c] (b);
		\vertex [above right = of c] (d);
		\vertex [below right = of c] (e);
		\diagram*{
		(a) -- [scalar] (c) -- [scalar] (b)
		(d) -- [scalar] (c) -- [scalar] (e)
		};
	\end{feynman}
	\end{tikzpicture},
\end{align}
where the solid line denotes the scalaron.
We have added the ``bar" to clarify that 
it is derived from the action~\eqref{eq:Einstein_HiggsR2}.
It is easily computed, and the $s$-wave part of the flavor-singlet 
elastic scattering amplitude is given by
\begin{align}
	\bar{a}_{\mathrm{tree}}^{(0)} = -\frac{N_s\left(1+6\xi\right)^2}{2304\pi\bar{\alpha}_1}
	\left[\frac{\bar{m}_s^2}{s-\bar{m}_s^2}+1\right],
	\label{eq:barred_amp}
\end{align}
to the leading order in $N_s$,
where the scalaron mass squared $\bar{m}_s^2$ is given by
\begin{align}
	\bar{m}_s^2 = \frac{\Mp^2}{12\bar{\alpha}_1},
\end{align}
and we have again omitted the contribution from the potential.
The factor $N_s$ again comes from 
the combinatory factor and the normalization of the state.
Thus, it coincides with 
the $s$-wave part of the self-healed amplitude
with $\bar{\alpha}_1 = \alpha_1$.
Remember that this identification is consistent with the beta function $\beta_{\alpha_1}$.
Based on this observation, we conclude that the self-healed amplitude
can be understood as the amplitude with the scalaron as the intermediate state.
It strongly supports that the Higgs inflation inevitably contains
the scalaron as a dynamical degree of freedom.

The physical meaning of ignoring the imaginary part 
in Eq.~\eqref{eq:self_heal_amp} is now clear.
When we compute the barred amplitude~\eqref{eq:barred_amp},
we have included only the tree-level contribution.
Once we include loop corrections, however,
the scalaron propagator develops an imaginary part
that corresponds to the decay rate.
In this sense, it is the tree-level approximation to ignore the imaginary part.
We can indeed verify that the imaginary part of Eq.~\eqref{eq:self_heal_amp}
is consistent with the decay rate of the scalaron.
Assuming that $\vert \ln(s/\mu_1^2)\vert \gg \pi$ and treating the logarithm as a constant,
the imaginary part shifts the position of the pole as
\begin{align}
	s = m_s^2 - i m_s \Gamma_s,
\end{align}
where we have defined
\begin{align}
	\Gamma_s \equiv -\frac{\pi m_s}{\ln(s/\mu_1^2)}
	= \frac{N_s\left(1+6\xi\right)^2}{192\pi}\frac{m_s^3}{\Mp^2}.
\end{align}
On the other hand, the decay rate of the scalaron is computed
from the action~\eqref{eq:Einstein_HiggsR2} as\footnote{
	The scalaron does not decay into a graviton pair
	although it linearly couples to $R$ 
	in the Jordan frame~\cite{Ema:2015dka,Ema:2016hlw}.
}
\begin{align}
	\bar{\Gamma}_s = \frac{N_s \left(1+6\xi\right)^2}{192\pi} \frac{\bar{m}_s^3}{\Mp^2}.
\end{align}
They coincide with each other under the identification $\alpha_1 = \bar{\alpha}_1$.
The condition $\vert \ln(s/\mu_1^2)\vert \gg \pi$ guarantees
that the scalaron has a narrow width.
Otherwise the perturbativity of the action~\eqref{eq:Einstein_HiggsR2} is lost.
It is also naturally expected that loop corrections induce a logarithmic dependence
of $\bar{\alpha}_1$ on $s$ as in Eq.~\eqref{eq:alpha1_self_heal}.
We leave an explicit confirmation of it as a future work.

In hindsight, the result may be naturally understood as follows.
It is the scalar loop contribution to the graviton vacuum polarization
that induces the $\alpha_1$-term.
The resummation of this contribution is equivalent
to dealing with the $\alpha_1$-term as a zero-th oder term in perturbation,
thus equivalent to treating the scalaron as a fundamental degree of freedom.
The above computation confirms this intuition.

Here we emphasize that the limit $\xi \gg 1$ and $N_s \gg 1$
is important since it justifies to include the $\alpha_1$-term to all orders while drop
other terms including the $\alpha_2$-term.
The $\alpha_2$-term is suppressed by $\xi$ compared to the $\alpha_1$-term,
while other (higher dimensional) terms are suppressed at least by $N_s$.
The large $\xi$ limit is probably good for $\xi = \mathcal{O}(10^4)$,
while the validity of the large $N_s$ limit could be a subject of discussion
for the Higgs inflation with $N_s = 4$.
We leave a detailed study on this point as a future work (see also Sec.~\ref{sec:summary}).

Before closing this section, 
let us clarify differences between our analysis and Ref.~\cite{Calmet:2013hia}.
The latter also discusses the self-healing mechanism of the Higgs inflation,
concluding that the unitarity is self-healed.
However, it does not discuss any physical interpretation of the self-healed amplitude.\footnote{
	Refs.~\cite{Calmet:2013hia,Calmet:2016fsr}
	rather claim that there is no physical pole in the $s$-wave sector,
	saying that the denominator is of the form $1 - sF_1(s)/2$ with $F_1(s) < 0$.
	We disagree for two reasons.
	First, their $F_1$ can be positive depending on $\mu_1$ 
	(or the parameter choice of the theory)
	since it is proportional to $\ln\left(s/\mu_1^2\right)$.
	Second, the amplitude has a pole even if $F_1$ is negative,
	although the corresponding particle (i.e. the scalaron) is tachyonic
	in such a case.
}
Our main point in this section is that the self-healed amplitude can be understood
as the amplitude with the scalaron as the intermediate state, 
thus to provide a physical interpretation to the self-healing mechanism
of the Higgs inflation.

\section{Equivalence of Jordan and Einstein frames}
\label{sec:equivalence}

So far we have discussed the self-healing mechanism and the emergence of the scalaron
in the Jordan frame, i.e. in the frame where the non-minimal coupling to the Ricci scalar is present.
It can be cast into the Einstein frame where the non-minimal coupling is absent
by the Weyl transformation.
Since people often discuss physics in the Einstein frame,
it is valuable to explicitly check that the same results are obtained in the Einstein frame.
For this purpose, we show the equivalence of the Jordan and Einstein frames 
of our results in this section.

We consider our action:
\begin{align}
	S = \int d^4x \sqrt{-g_J}\left[
	\frac{\Mp^2}{2}R_J
	+ \frac{1}{2}g^{\mu\nu}_J\partial_\mu \phi_i \partial_\nu \phi_i
	+ \frac{\xi}{2}R_J \phi_i^2
	\right]
	+ S_\mathrm{g.f.},
	\label{eq:action_Jordan}
\end{align}
where $S_\mathrm{g.f}$ is the gauge fixing term 
that is important in the following discussion,
and we have ignored the potential as usual.
We assign the subscripts $J$ and $E$ to the quantities in the Jordan frame and the Einstein frame,
respectively.
We perform the Weyl transformation as
\begin{align}
	g_{J\mu\nu} = \Omega^{-2} g_{E\mu\nu},
	~~~
	\Omega^2 = 1+ \frac{\xi \phi_i^2}{\Mp^2}.
\end{align}
The Ricci scalar is transformed as
\begin{align}
	R_J = \Omega^2 \left[ R_E + \frac{3}{2}g^{\mu\nu}_E 
	\partial_\mu \ln \Omega^2 \partial_{\nu} \ln \Omega^2
	- 3\DAlambert_E \ln \Omega^2\right].
\end{align}
As a result, the action in the Einstein frame is given by
\begin{align}
	S = \int d^4x \sqrt{-g_E}\left[
	\frac{\Mp^2}{2}R_E
	+ \frac{1}{2 \Omega^4}\left(\Omega^2 \delta_{ij} + \frac{6 \xi^2 \phi_i \phi_j}{\Mp^2}\right)
	g^{\mu\nu}_E\partial_\mu \phi_i \partial_\nu \phi_j
	\right]
	+ S_\mathrm{g.f.}.
	\label{eq:action_Einstein}
\end{align}
The metric of the scalar kinetic term is curved for $N_s \geq 2$, and hence 
$\xi$ is physical. 
The situation is different for $N_s = 1$, which is discussed separately in Sec.~\ref{sec:single}.

The Weyl transformation is merely a field redefinition, 
and physics is expected to be the same in both frames~\cite{George:2013iia,Postma:2014vaa,
Karam:2017zno}.
In this section, we explicitly check that this equivalence holds in our case.
In particular, we will see that the Feynman rules around $\phi_i = 0$ 
for 2-to-2 scattering are exactly the same
even at \textit{off-shell} level, once we properly take into account contributions from the gauge fixing term.
It enables us to see the equivalence parts by parts in Feynman diagrams.

\subsection{Jordan frame}
\label{subsec:Jordan}

First, we summarize the Feynman rules in the Jordan frame.
Note that all the diagrams of our interest are constructed from 
the \textit{off-shell} 2-to-2 scattering diagram,
\begin{align}
		\begin{tikzpicture} [baseline=(c)]
	\begin{feynman} [inline]
		\vertex (c);
		\vertex [above left = of c] (a);
		\vertex [below left = of c] (b);
		\vertex [right = of c] (d);
		\vertex [above right = of d] (e);
		\vertex [below right = of d] (f);
		\diagram*{
		(a) -- [scalar] (c) -- [scalar] (b)
		(c) -- [photon] (d)
		(e) -- [scalar] (d) -- [scalar] (f)
		};
	\end{feynman}
	\end{tikzpicture}.
\end{align}
We may take the de~Donder gauge in the Jordan frame, namely,
\begin{align}
	S_\mathrm{g.f.} 
	= \int d^4x \left[\left(\partial_{\nu}h_J^{\mu\nu} - \frac{1}{2}\partial^\mu h_J\right)
	\left(\partial^{\rho}h_{J\mu\rho}-\frac{1}{2}\partial_{\mu}h_J\right)\right],
\end{align}
at the quadratic order, where $g_{J\mu\nu} = \eta_{\mu\nu} + (2/\Mp) h_{J\mu\nu}$,
and $h_J \equiv \eta_{\mu\nu}h^{\mu\nu}_J$.
Then the Feynman rule for the above diagram is
\begin{align}
	\begin{tikzpicture} [baseline=(c)]
	\begin{feynman} [inline]
		\vertex (c);
		\vertex [label=\(p_1{,} i\), above left = of c] (a);
		\vertex [label=270:\(p_2{,} i\), below left = of c] (b);
		\vertex [right = of c] (d);
		\vertex [label=\(p_3{,} j\), above right = of d] (e);
		\vertex [label=270:\(p_4{,} j\), below right = of d] (f);
		\diagram*{
		(a) -- [scalar] (c) -- [scalar] (b)
		(c) -- [photon] (d)
		(e) -- [scalar] (d) -- [scalar] (f)
		};
	\end{feynman}
	\end{tikzpicture}
	&= -\frac{i}{2\Mp^2}\frac{1}{q^2}V_{\mu\nu}^{(0)}(p_1, p_2)
	P^{\mu\nu\rho\sigma}V^{(0)}_{\rho\sigma}(p_3,p_4)
	+ \frac{6i\xi^2}{\Mp^2}q^2 \nonumber \\
	&+ \frac{4i\xi}{\Mp^2}\left[\left(p_1\cdot p_2\right) + \left(p_3\cdot p_4\right)
	- \frac{\left(p_1\cdot q\right)\left(p_2 \cdot q\right) + \left(p_3 \cdot q\right)\left(p_4 \cdot q\right)}{q^2}
	\right],
	\label{eq:Feynman_Jordan}
\end{align}
where $q \equiv p_1 + p_2 = p_3 + p_4$, we take the external momenta $p_1, p_2, p_3$ and $p_4$
to be off-shell, and
\begin{align}
	V^{(0)}_{\mu\nu}(p,q) &\equiv \left(p\cdot q\right) \eta_{\mu\nu} - \left(p_\mu q_\nu + p_\nu q_\mu\right), \\
	P_{\mu\nu\rho\sigma} & \equiv \eta_{\mu\rho}\eta_{\nu\sigma} + \eta_{\mu\sigma}\eta_{\nu\rho}
	- \eta_{\mu\nu}\eta_{\rho\sigma}.
\end{align}
The function $V^{(0)}_{\mu\nu}$ corresponds to the scalar-scalar-graviton vertex for $\xi = 0$.
See App.~\ref{app:details} for more details.
Below we see that the same off-shell Feynman rule is obtained in the Einstein frame
once we properly take into account contributions from the gauge fixing term.

\subsection{Einstein frame}
\label{subsec:Einstein}
Now we derive the Feynman rules for off-shell 2-to-2 scattering diagrams in the Einstein frame.
The action~\eqref{eq:action_Einstein} contains scalar four-point vertices,
\begin{align}
	S \ni \int d^4x \left[-\frac{\xi \phi_j^2}{2\Mp^2} \partial^\mu \phi_i \partial_\mu \phi_i 
	+ \frac{3\xi^2}{4\Mp^2} \left(\partial^\mu \phi_i^2\right) \left(\partial_\mu \phi_j^2\right) \right],
\end{align}
as well as the standard scalar-scalar-graviton vertex.
There are additional contributions from the gauge fixing term.
The gravitons in the Jordan and Einstein frames are related as
\begin{align}
	h_{J\mu\nu} &= \frac{1}{\Omega^2}\left(\frac{\Mp}{2}\eta_{\mu\nu} + h_{E\mu\nu}\right) 
	- \frac{\Mp}{2}\eta_{\mu\nu} 
	\nonumber \\
	&= h_{E\mu\nu} - \frac{\xi \phi_i^2}{2\Mp}\eta_{\mu\nu} + ...,
\end{align}
and hence the de~Donder gauge in the Jordan frame results in additional vertices in the Einstein frame:
\begin{align}
	S_{\mathrm{g.f.}} \ni \int d^4 x 
	\left[ 
	-\frac{\xi}{\Mp}\left(\partial^\mu \partial^\nu \phi_i^2 - \frac{1}{2}\eta^{\mu\nu} \partial^2 \phi_i^2\right)
	h_{E\mu\nu} + \frac{\xi^2}{4\Mp^2}\left(\partial^\mu\phi_i^2\right)\left(\partial_\mu \phi_j^2\right)
	\right].
	\label{eq:scalar_four}
\end{align}
The Feynman rules are thus given by
\begin{align}
	\begin{tikzpicture} [baseline=(c)]
	\begin{feynman} [inline]
		\vertex (c);
		\vertex [label=\(p_1{,} i\), above left = of c] (a);
		\vertex [label=270:\(p_2{,} i\), below left = of c] (b);
		\vertex [right = of c] (d);
		\diagram*{
		(a) -- [scalar] (c)
		(b) -- [scalar] (c)
		(c) -- [photon] (d)
		};
	\end{feynman}
	\end{tikzpicture}
	= -\frac{i}{\Mp}V_{\mu\nu}^{(0)}(p_1, p_2) + \frac{2i\xi}{\Mp}
	\left(q_\mu q_\nu - \frac{\eta_{\mu\nu}}{2}q^2\right),
\end{align}
for the scalar-scalar-graviton vertex,
and
\begin{align}
	\begin{tikzpicture} [baseline=(c)]
	\begin{feynman} [inline]
		\vertex (c);
		\vertex [label=\(p_1{,} i\), above left = of c] (a);
		\vertex [label=270:\(p_2{,} i\), below left = of c] (b);
		\vertex [label=\(p_3{,} j\), above right = of c] (d);
		\vertex [label=270:\(p_4{,} j\), below right = of c] (e);
		\diagram*{
		(a) -- [scalar] (c) -- [scalar] (b)
		(d) -- [scalar] (c) -- [scalar] (e)
		};
	\end{feynman}
	\end{tikzpicture}
	= \frac{2i\xi}{\Mp^2}\left( \left(p_1\cdot p_2\right) + \left( p_3 \cdot p_4 \right) \right)
	+ \frac{8i\xi^2}{\Mp^2}q^2,
\end{align}
for the scalar four-point vertex with $i \neq j$, where again $q \equiv p_1 + p_2$.
We now obtain the Feynman rule for the off-shell 2-to-2 scattering diagrams as
\begin{align}
	\begin{tikzpicture} [baseline=(c)]
	\begin{feynman} [inline]
		\vertex (c);
		\vertex [label=\(p_1{,} i\), above left = of c] (a);
		\vertex [label=270:\(p_2{,} i\), below left = of c] (b);
		\vertex [right = of c] (d);
		\vertex [label=\(p_3{,} j\), above right = of d] (e);
		\vertex [label=270:\(p_4{,} j\), below right = of d] (f);
		\diagram*{
		(a) -- [scalar] (c) -- [scalar] (b)
		(c) -- [photon] (d)
		(e) -- [scalar] (d) -- [scalar] (f)
		};
	\end{feynman}
	\end{tikzpicture}	
	+
	\begin{tikzpicture} [baseline=(c)]
	\begin{feynman} [inline]
		\vertex (c);
		\vertex [label=\(p_1{,} i\), above left = of c] (a);
		\vertex [label=270:\(p_2{,} i\), below left = of c] (b);
		\vertex [label=\(p_3{,} j\), above right = of c] (d);
		\vertex [label=270:\(p_4{,} j\), below right = of c] (e);
		\diagram*{
		(a) -- [scalar] (c) -- [scalar] (b)
		(d) -- [scalar] (c) -- [scalar] (e)
		};
	\end{feynman}
	\end{tikzpicture}
	&= -\frac{i}{2\Mp^2}\frac{1}{q^2}V_{\mu\nu}^{(0)}(p_1, p_2)
	P^{\mu\nu\rho\sigma}V^{(0)}_{\rho\sigma}(p_3,p_4)
	+ \frac{6i\xi^2}{\Mp^2}q^2 \nonumber \\
	&+ \frac{4i\xi}{\Mp^2}\left[\left(p_1\cdot p_2\right) + \left(p_3\cdot p_4\right)
	- \frac{\left(p_1\cdot q\right)\left(p_2 \cdot q\right) + \left(p_3 \cdot q\right)\left(p_4 \cdot q\right)}{q^2}
	\right],
	\label{eq:Feynman_Einstein}
\end{align}
for $i\neq j$, 
which is exactly the same as the Feynman rule~\eqref{eq:Feynman_Jordan} 
in the Jordan frame.
We can show the equivalence for $i = j$ in a similar manner
once we include not only the $s$-channel but also the $t$- and $u$-channels, 
as they all contribute for $i = j$.
The equivalence of the Feynman rules associated with the counter terms
can be confirmed in the same way.\footnote{
	To the order of our interest,
	the $\alpha_1$-term in the Jordan frame results in a scalar-scalar-graviton vertex
	and a scalar four-point vertex in addition to the graviton quadratic term~\eqref{eq:grav_quad_alpha1}
	in the Einstein frame.
	The $\alpha_2$-term in the Jordan frame does not induce vertices involving the scalar fields
	in the Einstein frame.
} Thus, we have shown the equivalence of the Jordan and Einstein frames 
at the diagrammatic level. The equivalence of our argument,
especially the self-healing mechanism, 
follows straightforwardly.

In the above argument we have carefully taken into account 
the vertices arising from the gauge fixing term.
The choice of the gauge fixing term is unphysical,
and hence we will obtain the same result even if we do not include those vertices
as far as physical quantities are concerned.\footnote{
	For instance, we can start with the de~Donder gauge in the Einstein frame,
	and compute quantities such as the scalar loop to the graviton vacuum polarization.
	We have checked that it reproduces the same result as expected.
}
Nevertheless, we feel it meaningful to treat the gauge fixing term carefully,
since it enables us to see the equivalence
not only at the on-shell level, but also at the off-shell (i.e. Feynman diagrammatic) level.
For instance, we now easily see that the self-healed amplitude~\eqref{eq:sum}
in the Jordan frame roughly corresponds to
\begin{align}
	A_{\mathrm{dressed}}^{(ii\rightarrow jj)} &=
	\begin{tikzpicture}[baseline=(c)]
	\begin{feynman}[inline]
		\vertex (c);
		\vertex [above left = of c] (a);
		\vertex [below left = of c] (b);
		\vertex [above right = of c] (e);
		\vertex [below right = of c] (f);
		\diagram*{
		(a) -- [scalar] (c)
		(b) -- [scalar] (c)
		(c) -- [scalar] (e)
		(c) -- [scalar] (f)
		};
	\end{feynman}
	\end{tikzpicture}
	+
	\begin{tikzpicture}[baseline=(c)]
	\begin{feynman}[inline]
		\vertex (c);
		\vertex [above left = of c] (a);
		\vertex [below left = of c] (b);
		\vertex [right = of c] (d);
		\vertex [above right = of d] (g);
		\vertex [below right = of d] (h);
		\diagram*{
		(a) -- [scalar] (c) -- [scalar] (b)
		(c) -- [scalar, half left] (d) -- [scalar, half left] (c)
		(g) -- [scalar] (d) -- [scalar] (h) 
		};
	\end{feynman}
	\end{tikzpicture}
	+
	\begin{tikzpicture}[baseline=(c)]
	\begin{feynman}[inline]
		\vertex (c);
		\vertex [above left = of c] (a);
		\vertex [below left = of c] (b);
		\vertex [right = of c] (d);
		\vertex [right = of d] (e);
		\vertex [above right = of e] (f);
		\vertex [below right = of e] (g);
		\diagram*{
		(a) -- [scalar] (c) -- [scalar] (b)
		(c) -- [scalar, half left] (d) -- [scalar, half left] (c)
		(e) -- [scalar, half left] (d) -- [scalar, half left] (e)
		(f) -- [scalar] (e) -- [scalar] (g) 
		};
	\end{feynman}
	\end{tikzpicture}
	+ \cdots,
	\label{eq:sum_Einstein}
\end{align}
in the Einstein frame.
In other words, the self-healing mechanism in the Einstein frame
comes dominantly from the non-renormalizable interaction
in the scalar kinetic terms.
This observation has an interesting implication,
which we will briefly discuss in Sec.~\ref{sec:summary}.

\section{Single field case}
\label{sec:single}

So far we have focused on the case $N_s \geq 2$. 
The single field case $N_s = 1$ is tricky
and is discussed separately in this section.
Remember that $N_s = 4$ in the Higgs inflation,
and hence the subtlety discussed in the following does not apply to that case.
We ignore the potential first, 
and briefly comment on effects of the potential in the end.

It is known that the non-minimal coupling $\xi$ is unphysical
for $N_s = 1$ once we ignore the potential, as emphasized in Ref.~\cite{Hertzberg:2010dc}.
It is best described by the Weyl transformation.
We consider the single field action with the non-minimal coupling:
\begin{align}
	S = \int d^4x \sqrt{-g_J}\left[
	\frac{\Mp^2}{2}R_J
	+ \frac{1}{2}g^{\mu\nu}_J\partial_\mu \phi \partial_\nu \phi
	+ \frac{\xi}{2}R_J \phi^2
	\right].
	\label{eq:action_Jordan_single}
\end{align}
We can move to the Einstein frame in the same way to obtain Eq.~\eqref{eq:action_Einstein}
from Eq.~\eqref{eq:action_Jordan}.
The action in the Einstein frame is
\begin{align}
	S = \int d^4x \sqrt{-g_E}\left[
	\frac{\Mp^2}{2}R_E
	+ \frac{1}{2 \Omega^4}\left(\Omega^2 + \frac{6 \xi^2 \phi^2}{\Mp^2}\right)
	g^{\mu\nu}_E\partial_\mu \phi \partial_\nu \phi
	\right].
	\label{eq:action_Einstein_single}
\end{align}
Now we redefine the scalar field as
\begin{align}
	d\chi = \frac{d\phi}{\Omega^2}\sqrt{\Omega^2 + \frac{6\xi^2 \phi^2}{\Mp^2}},
	\label{eq:field_redef_single}
\end{align}
so that the action is recast into the Einstein-Hilbert term and a canonically normalized scalar field.
The non-minimal coupling $\xi$ can be erased by the field redefinition, and hence is unphysical.
The crucial difference between Eq.~\eqref{eq:action_Einstein} and~\eqref{eq:action_Einstein_single}
is that the target space of the scalar fields is curved for $N_s \geq 2$, 
while it is not curved for $N_s = 1$.
The curvature depends on $\xi$ for $N_s \geq 2$,
indicating that $\xi$ is physical as the curvature is invariant under field redefinition
(that corresponds to coordinate transformation of the target space).

It is possible to check explicitly that physical quantities
such as scattering amplitudes do not depend on $\xi$.
At the tree-level, the $s$-channel scattering amplitude of $\phi \phi \rightarrow \phi \phi$ 
is given by Eq.~\eqref{eq:amp_tree}, i.e., 
\begin{align}
	A_\mathrm{tree}^{(s)} 
	= \frac{1}{\Mp^2s}\left[\frac{\left(1+6\xi\right)^2}{6} s^2 - \left(\frac{s^2}{6}-tu\right)\right].
\end{align}
In the present case, the $t$- and $u$-channels also contribute to the scattering amplitude,
and hence the tree-level amplitude is given by
\begin{align}
	A_\mathrm{tree} = A_\mathrm{tree}^{(s)} + A_\mathrm{tree}^{(t)} + A_\mathrm{tree}^{(u)}
	=
	\frac{1}{\Mp^2}\left[ \frac{tu}{s} + \frac{us}{t} + \frac{st}{u} \right].
\end{align}
In particular, the $\xi$-dependent part drops since the Mandelstam variables satisfy $s + t + u = 0$.
At the one-loop level, it is probably easiest to work with the de~Donder gauge in the Einstein frame
(that results in additional vertices in the Jordan frame,
in a similar way as discussed in Sec.~\ref{sec:equivalence}).
For instance, we can easily see in the Einstein frame (before the field redefinition) 
that the following one-loop diagram vanishes:
\begin{align}
	iA_{\mathrm{1\mathchar`-loop}}^{(s)} =
	\begin{tikzpicture}[baseline=(c)]
	\begin{feynman}[inline]
		\vertex (c);
		\vertex [above left = of c] (a);
		\vertex [below left = of c] (b);
		\vertex [right = of c] (d);
		\vertex [above right = of d] (g);
		\vertex [below right = of d] (h);
		\diagram*{
		(a) -- [scalar] (c) -- [scalar] (b)
		(c) -- [scalar, half left] (d) -- [scalar, half left] (c)
		(g) -- [scalar] (d) -- [scalar] (h) 
		};
	\end{feynman}
	\end{tikzpicture}
	= 0.
	\label{eq:one_loop_Einstein_single}
\end{align}
In a similar way as in Sec.~\ref{sec:equivalence},
the scalar four-point vertex in the Einstein frame 
is equivalent at the off-shell level to the diagrams in the Jordan frame as
\begin{align}
	\begin{tikzpicture} [baseline=(c)]
	\begin{feynman} [inline]
		\vertex (c);
		\vertex [above left = of c] (a);
		\vertex [below left = of c] (b);
		\vertex [above right = of c] (d);
		\vertex [below right = of c] (e);
		\diagram*{
		(a) -- [scalar] (c) -- [scalar] (b)
		(d) -- [scalar] (c) -- [scalar] (e)
		};
	\end{feynman}
	\end{tikzpicture}
	\sim
	\begin{tikzpicture} [baseline=(c)]
	\begin{feynman} [inline]
		\vertex (c);
		\vertex [above left = of c] (a);
		\vertex [below left = of c] (b);
		\vertex [right = of c] (d);
		\vertex [above right = of d] (e);
		\vertex [below right = of d] (f);
		\diagram*{
		(a) -- [scalar] (c) -- [scalar] (b)
		(c) -- [photon] (d)
		(e) -- [scalar] (d) -- [scalar] (f)
		};
	\end{feynman}
	\end{tikzpicture}	
	+
	(s \leftrightarrow t)
	+
	(s \leftrightarrow u)
	+
	\begin{tikzpicture} [baseline=(c)]
	\begin{feynman} [inline]
		\vertex (c);
		\vertex [above left = of c] (a);
		\vertex [below left = of c] (b);
		\vertex [above right = of c] (d);
		\vertex [below right = of c] (e);
		\diagram*{
		(a) -- [scalar] (c) -- [scalar] (b)
		(d) -- [scalar] (c) -- [scalar] (e)
		};
	\end{feynman}
	\end{tikzpicture},
\end{align}
where we have used the similarity since the right hand side includes 
the contributions from the spin-2 part of the graviton as well.\footnote{
	The de~Donder gauge in one frame generates
	a scalar four-point vertex in the other frame,
	as in Eq.~\eqref{eq:scalar_four}.
}
It is thus clear that the one-loop diagram~\eqref{eq:one_loop_Einstein_single}
in the Einstein frame corresponds to many diagrams including the vacuum polarization,
the vertex correction and the box diagrams in the Jordan frame.
It means that the $\xi$-dependence cancels out among these diagrams in the Jordan frame.
The diagrams with the scalar loop depend on $N_s$ differently
from the other diagrams, and hence this cancellation holds only for $N_s = 1$.\footnote{
 	The beta functions in Sec.~\ref{sec:RGE} take into account only the scalar loop.
	While there may be additional contributions from 
	the vertex correction and the box diagrams,
	the argument in Sec.~\ref{sec:RGE} is intact 
	since the cancellation does not occur for $N_s \geq 2$.
	Also we can ignore these terms in Sec.~\ref{sec:self_healing} as
	they are sub-leading in the large $N_s$ expansion.
 }

It follows
that the cut-off scale of the theory with $N_s = 1$ does not depend on $\xi$, 
and hence the self-healing mechanism is not at work (at least below the Planck scale).
Thus, the dynamical emergence of the scalaron
does not happen in the single field case without the potential.
Once we include the potential, the story could be totally different.
The field redefinition~\eqref{eq:field_redef_single} results in
higher dimensional operators in the potential,
whose cut-off scale does depend on $\xi$~\cite{Hertzberg:2010dc}.
We emphasize that this case 
requires a separate discussion from the case $N_s \geq 2$.
At least, there is no scalar four-point derivative interaction in the Einstein frame
(after field redefinition), 
and hence the self-healing mechanism (if at work) will not be 
in the form of Eq.~\eqref{eq:sum_Einstein}.
Moreover, we may expect that 
a particle that emerges due to the self-healing mechanism (if any)
is not the scalaron, since the scalaron may not cure
the low cut-off scale associated with the higher dimensional operators
in the potential sector.
A detailed study on this case, albeit definitely interesting, is beyond the scope of this paper.

\section{Summary and discussions}
\label{sec:summary}

In this paper, we have argued that a light scalaron
dynamically emerges if there is a sizable non-minimal coupling between scalar fields
and the Ricci scalar $\xi \phi_i^2 R$ as in the Higgs inflation model,
as long as the number of the scalar fields $N_s$ satisfies $N_s \geq 2$.
This claim is based on two (closely related) observations. 
First, the $R^2$-term necessarily emerges due to the RG running
whose coefficient is generically of $\mathcal{O}(\xi^2)$ for $\xi \gg 1$.
Since the coefficient of the $R^2$-term is inversely proportional 
to the scalaron mass squared, it implies that a light scalaron inevitably exists in the theory
(see also Refs.~\cite{Salvio:2015kka,Calmet:2016fsr,Ema:2017rqn,Ghilencea:2018rqg}).
Second, we have seen that the 2-to-2 scattering amplitude $\phi_i \phi_i \rightarrow \phi_j \phi_j$
with $i \neq j$ develops a pole structure after resumming over the diagrams that are leading order
in the large $N_s$ expansion.
We have explicitly checked that the resultant scattering amplitude is equivalent
to the amplitude with the scalaron in the intermediate state.
The resummation heals the unitarity of the Higgs inflation, 
called the self-healing mechanism in Refs.~\cite{Aydemir:2012nz,Calmet:2013hia},
and hence our result identifies the self-healing mechanism 
as the dynamical emergence of the scalaron.
We have confirmed that our results do not depend on whether we work in the Jordan frame 
or in the Einstein frame.

Two implications immediately follow.
First, the Higgs inflation is actually a two-field inflationary model,
the (radial component of the) Higgs and the scalaron.
In particular, the inflaton is in general an admixture of the Higgs and the scalaron.
The inflationary dynamics of this theory is discussed in 
App.~\ref{app:HiggsR2} and references therein.
Second, contrary to the common wisdom, 
the Higgs inflation does not suffer from the unitarity issue
since the self-healing mechanism is at work.
It is nothing but the fact that the scalaron pushes up the cut-off scale to the Planck scale,
as pointed out in Refs.~\cite{Ema:2017rqn,Gorbunov:2018llf}.
Thus, we can follow all the dynamics from the inflation 
till the end of the reheating within the validity of the theory.

There are several other points that are not discussed in detail in this paper.
Below we list some of them before ending this paper.

\subsubsection*{Self-healing in other inflationary models}

In Sec.~\ref{sec:equivalence}, we see that the self-healing mechanism of the Higgs inflation
is caused by the higher dimensional operators in the scalar kinetic sector
in the Einstein frame (see Eq.~\eqref{eq:sum_Einstein}).
There are other inflationary models that have non-trivial scalar kinetic terms
(or a curved target space),
such as the running kinetic model~\cite{Takahashi:2010ky,Nakayama:2010kt},
the $\alpha$-attractor model~\cite{Kallosh:2013yoa},
and the Higgs-dilaton model~\cite{GarciaBellido:2011de}.
The Palatini Higgs inflation model~\cite{Bauer:2010jg,Rasanen:2017ivk,Antoniadis:2018ywb,
Antoniadis:2018yfq,Tenkanen:2019jiq} 
could also be within this class of models in the Einstein frame, among probably many others.
Since these non-trivial kinetic terms are interpreted as higher dimensional operators 
once expanded around the vacuum,
a similar self-healing mechanism may be at work
and a new degree of freedom may show up dynamically in these models.
This new degree of freedom, if any, is not necessarily be the scalaron.
In particular, it is possible in some models 
that the new degree of freedom is pathological, either ghost-like or tachyonic,
that may spoil the inflationary prediction.
Also it is possible that there appear more than one new degrees of freedom,
which might drastically modify the inflationary prediction.
These points will be clarified
once we study pole structures of scattering amplitudes
resulting from resummation similar to Eq.~\eqref{eq:sum_Einstein}.
For instance, the position of the pole tells whether a particle
is tachyonic or not, while the sign of the residue does whether it
is ghost-like or not.\footnote{
	The spin-2 ($d$-wave) part of the self-healed amplitude in our theory
	also develops a pole.
	Its residue has an opposite sign from that of the spin-0 ($s$-wave) part,
	as long as the poles are not tachyonic.
	Since the unitarity requires that the residues are expanded 
	by the Legendre polynomials whose coefficients have a definite sign, 
	it indicates that a ghost exists in the spin-2 sector.
	The position of the pole is consistent with Eq.~\eqref{eq:mass_spin2}.
} We believe that it is of great importance to re-analyze 
the other inflationary models
from this point of view, which will be done elsewhere.

\subsubsection*{Running of scalaron mass}

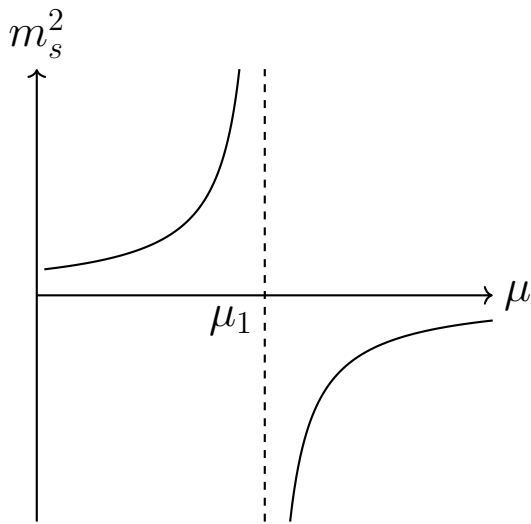
\begin{figure}[t]
	\centering
	\begin{tikzpicture} [domain=0:6, samples=100, thick]
	\draw (0,0) node[below left]{};
	\draw[->] (0,0)--(6,0) node[right] {\Large $\mu$};
	\draw[->] (0,-3)--(0,3) node[above] {\Large $m_s^2$};
	\draw[domain=0.1:2.6666] plot(\x, {1/(3-\x)});
	\draw[domain=3.3333:6] plot(\x, {1/(3-\x)});
	\draw[dashed] (3,-3)--(3,3);
	\draw (3,0) node[below left]{\Large $\mu_1$};
	\end{tikzpicture}
	\caption{\small Schematic picture of the RG running of the scalaron mass squared $m_s^2$.}
	\label{fig:RGE}
\end{figure}

The running of the scalaron mass squared shows a peculiar behavior.
The (scalar loop contribution to the) RG equation of $\alpha_1$ is given by
\begin{align}
	\beta_{\alpha_1} &\equiv \frac{d\alpha_1}{d\ln \mu} = 
	-\frac{N_s}{1152\pi^2}\left(1+6\xi\right)^2.
\end{align}
Since the scalaron mass squared $m_s^2$ in terms of $\alpha_1$ is
\begin{align}
	m_s^2 = \frac{\Mp^2}{12\alpha_1},
\end{align}
its RG running is schematically given in Fig.~\ref{fig:RGE}.\footnote{
	It crucially depends on the sign of the beta function.
	See the footnote~\ref{ft:sign_beta}.
} An interesting feature is that its mass squared diverges at the energy scale $\mu_1$
at which $\alpha_1$ vanishes,
and becomes tachyonic above $\mu_1$.
It might be hard to imagine that the low energy theory is healthy above $\mu_1$.
An easy way to avoid such a situation is to assume that
UV completion of the gravity provides $\alpha_1 > 0$ as the boundary condition,
i.e. $\mu_1 > \Lambda_\mathrm{UV}$ where $\Lambda_\mathrm{UV}$
is the scale at which the UV completion comes into play.
We may further require $\ln(\mu_1^2/\Lambda_\mathrm{UV}^2) \gg \pi$
to guarantee the perturbativity of the action~\eqref{eq:Einstein_HiggsR2}.

The RG equation indicates that it has some unease to assume that 
the scalaron is so heavy at the energy scale of the Higgs inflation (which we call $\mu_\mathrm{HI}$)
that it does not affect the inflationary dynamics.
If we assume $\alpha_1 \ll \xi^2$ at $\mu_\mathrm{HI}$,
the scalaron becomes tachyonic just above $\mu_\mathrm{HI}$.
It probably requires UV completion just above $\mu_\mathrm{HI}$ that is
significantly lower than the Planck scale.
Also the scalaron has to be taken into account after inflation such as reheating
even if it is heavy during inflation.

\subsubsection*{Large $N_s$ expansion}

In Sec.~\ref{sec:self_healing}, we have discussed the self-healing mechanism
relying on the large $N_s$ limit.
Based on the discussion in Sec.~\ref{sec:RGE}, however, it is natural to expect 
that the dynamical emergence of the scalaron is at work 
for finite $N_s$ as well.
A first step to check this expectation
may be to include next-to-leading order (NLO) terms in the large $N_s$ expansion.
At the NLO level, one is probably required to include diagrams 
with one graviton or ghost inside loops,
and diagonal parts of the scattering amplitude between the flavor-singlet states.
Computation will be of course more tedious, but it is nevertheless valuable to see
whether the self-healing mechanism works or not at the NLO level.
We leave it as a future work.

\subsubsection*{Amplitudes around $\phi_i \neq 0$}

In this paper we have focused on the scattering amplitudes around the vacuum $\phi_i = 0$.
However, the self-healing mechanism of the Higgs inflation is discussed also around 
the finite Higgs field value $\phi_i \neq 0$ in Ref.~\cite{Calmet:2013hia},
which concludes that it works in the same way for $\phi_i \neq 0$.
Thus we expect that our discussion in this paper
can be extended to the amplitudes around $\phi_i \neq 0$ as well.
It probably requires some cares including the prescription issue
(see e.g. Refs.~\cite{
Bezrukov:2008ej,Bezrukov:2009db,
George:2013iia,Postma:2014vaa,Falls:2018olk}
and references therein)
to see the equivalence between the Jordan and Einstein frames in this case.

\section*{Acknowledgement}
YE thanks Kyohei Mukaida for valuable discussions and motivating him to write this paper.
The Feynman diagrams in this paper are generated by \texttt{TikZ-Feynman}~\cite{Ellis:2016jkw}.

\appendix

\section{Computational details}
\label{app:details}
In this appendix, for the sake of clarity and completeness, 
we give our derivation of 
the scattering amplitudes and the RG equations
in detail.

\subsection{Preliminary}

\subsubsection*{Gravity sector}
Here we summarize our convention in this paper.
We use the almost-minus convention for the metric
(denoted as $g_{\mu\nu}$).
In particular, the flat spacetime metric is given by
\begin{align}
	\eta_{\mu\nu} = \mathrm{diag}\left(+1, -1, -1, -1\right).
\end{align}
The Christoffel symbol is given by
\begin{align}
	{\Gamma^\rho}_{\mu\nu} = \frac{g^{\rho\sigma}}{2}
	\left(g_{\mu \sigma, \nu} + g_{\nu \sigma, \mu} - g_{\mu\nu, \sigma}\right),
\end{align}
where $g_{\mu\nu, \rho} \equiv \partial_\rho g_{\mu\nu}$.
The sign convention for the Ricci tensor is chosen such that
\begin{align}
	R_{\mu\rho} = {\Gamma^\nu}_{\nu\rho,\mu} - {\Gamma^\nu}_{\mu\rho,\nu}
	+{\Gamma^\alpha}_{\nu\rho}{\Gamma^\nu}_{\alpha\mu}
	-{\Gamma^\alpha}_{\mu\rho}{\Gamma^\nu}_{\alpha\nu},
\end{align}
and the Ricci scalar is given by
\begin{align}
	R = g^{\mu\nu}R_{\mu\nu}.
\end{align}
It follows from these conventions that the variations are given by
\begin{align}
	\delta \sqrt{-g} &= -\frac{\sqrt{-g}}{2}g_{\mu\nu}\delta g^{\mu\nu}, \\
	\delta R &= R_{\mu\nu}\delta g^{\mu\nu} - g_{\mu\nu}  \DAlambert \delta g^{\mu\nu}
	+ \nabla_\mu \nabla_\nu \delta g^{\mu\nu},
\end{align}
where $g \equiv \mathrm{det}\left(g_{\mu\nu}\right)$, $\nabla_\mu$ is the covariant derivative
and $\DAlambert = \nabla^\alpha \nabla_\alpha$.

\subsubsection*{Energy momentum tensor}
In this paper we focus on the following action:\footnote{
The sign of the Einstein Hilbert action is fixed once the sign convention of the Ricci scalar is fixed.
}
\begin{align}
	S = \int d^4x \sqrt{-g}\left[
	\frac{\Mp^2}{2}R
	+ \frac{1}{2}g^{\mu\nu}\partial_\mu \phi_i \partial_\nu \phi_i
	+ \frac{\xi}{2}R\phi_i^2
	- V(\phi_i)
	\right],
	\label{eq:action_app}
\end{align}
where $i = 1, 2, ..., N_s$ is the flavor index ($N_s = 4$ in the Higgs inflation).
With this definition, $\xi = -1/6$ corresponds to the conformal coupling.
The potential $V(\phi_i)$ is not important in our discussion,
and is ignored henceforth.
The energy stress tensor is defined as
\begin{align}
	T_{\mu\nu} = -\frac{2}{\sqrt{-g}}\frac{\delta S_m}{\delta g^{\mu\nu}},
\end{align}
where $S_m$ is the matter part of the action.
It is given in our case as
\begin{align}
	T_{\mu\nu} =& \frac{g_{\mu\nu}}{2}\left(1+4\xi\right)g^{\alpha\beta}\partial_\alpha \phi_i \partial_\beta \phi_i
	-\left(1+2\xi\right)\partial_\mu \phi_i \partial_\nu \phi_i \nonumber \\
	&+ 2\xi\left(g_{\mu\nu}\phi_i \DAlambert \phi_i - \phi_i \nabla_\mu \nabla_\nu \phi_i \right)
	- \xi\left(R_{\mu\nu}-\frac{g_{\mu\nu}}{2}R\right)\phi_i\phi_i.
\end{align}
In the flat spacetime it reduces to
\begin{align}
	T_{\mu\nu}^{(\mathrm{flat})} 
	=& \frac{\eta_{\mu\nu}}{2}\left(1+4\xi\right)\partial^\alpha \phi_i \partial_\alpha \phi_i
	-\left(1+2\xi\right)\partial_\mu \phi_i \partial_\nu \phi_i 
	+ 2\xi\left(\eta_{\mu\nu}\phi_i \partial^2 \phi_i - \phi_i \partial_\mu \partial_\nu \phi_i \right).
\end{align}

\subsection{Feynman rules}
Here we summarize the Feynman rules in the flat spacetime derived from the action~\eqref{eq:action}.
The ghost degree of freedom associated with the gauge fixing is irrelevant for our analysis,
and hence we ignore it in the following.

\subsubsection*{Propagator}
The scalar propagator is given by
\begin{align}
	\begin{tikzpicture} [baseline=(a)]
	\begin{feynman}
		\vertex (a);
		\vertex [right= of a] (b);
		\diagram*{
		(a) -- [scalar] (b)
		};
	\end{feynman}
	\end{tikzpicture}
	~= \frac{i}{p^2 + i0}.
	\label{eq:scalar_prop}
\end{align}
We expand the metric around the flat spacetime as
\begin{align}
	g_{\mu\nu} = \eta_{\mu\nu} + \frac{2}{\Mp}h_{\mu\nu}.
\end{align}
We take the de~Donder gauge in the Jordan frame in this appendix, given by
\begin{align}
	S_\mathrm{g.f.} 
	= \int d^4x \left[\left(\partial_{\nu}h^{\mu\nu} - \frac{1}{2}\partial^\mu h\right)
	\left(\partial^{\rho}h_{\mu\rho}-\frac{1}{2}\partial_{\mu}h\right)\right],
\end{align}
at the quadratic order, and hence the graviton propagator is given by
\begin{align}
	\begin{tikzpicture} [baseline=(a)]
	\begin{feynman}
		\vertex [label=270:\(\mu\nu\)] (a);
		\vertex [label=270:\(\rho\sigma\), right= of a] (b);
		\diagram*{
		(a) -- [photon] (b)
		};
	\end{feynman}
	\end{tikzpicture}
	= \frac{i}{2p^2 + i0}
	P_{\mu\nu\rho\sigma},
	\label{eq:graviton_prop}
\end{align}
where
\begin{align}
	P_{\mu\nu\rho\sigma} \equiv
	\eta_{\mu\rho}\eta_{\nu\sigma} + \eta_{\mu\sigma}\eta_{\nu\rho}
	- \eta_{\mu\nu}\eta_{\rho\sigma}.
\end{align}

\subsubsection*{Scalar-scalar-graviton interaction}
The interaction between the scalar fields and the graviton is given by
\begin{align}
	S_\mathrm{int} = \frac{1}{\Mp}\int d^4x \sqrt{-g}\,h^{\mu\nu}T_{\mu\nu}.
\end{align}
From this expression, the scalar-scalar-graviton interaction reads
\begin{align}
	\begin{tikzpicture} [baseline=(c)]
	\begin{feynman} [inline]
		\vertex (c);
		\vertex [label=\(p_1{,} i\), above left = of c] (a);
		\vertex [label=270:\(p_2{,} j\), below left = of c] (b);
		\vertex [label=270:\(\mu\nu\), right = of c] (d);
		\diagram*{
		(a) -- [scalar] (c)
		(b) -- [scalar] (c)
		(c) -- [photon] (d)
		};
	\end{feynman}
	\end{tikzpicture}
	= -\frac{i \delta_{ij}}{\Mp} V_{\mu\nu}\left(p_1, p_2\right),
	\label{eq:scalar_scalar_graviton}
\end{align}
where the vertex function is given by
\begin{align}
	V_{\mu\nu}\left(p_1, p_2\right) =& 
	\left(1+4\xi\right)\left(p_1 \cdot p_2\right)\eta_{\mu\nu} 
	- \left(1+2\xi\right)\left(p_{1\mu}p_{2\nu} + p_{1\nu}p_{2\mu}\right) \nonumber \\
	&+2\xi \left[ \left(p_1^2 + p_2^2\right)\eta_{\mu\nu} - 
	\left(p_{1\mu}p_{1\nu} + p_{2\mu}p_{2\nu}\right) \right],
\end{align}
and the momenta are both outgoing or both incoming.\footnote{
	The factor two from the exchange of the scalars is included
	in this rule.
} Note that the first term in the second line does not vanish for off-shell momenta.
	
\subsubsection*{Counter terms}

As we will see in the following, we need to include counter terms to renormalize divergences
from the scalar one-loop diagram. Thus we include the following terms in our action:
\begin{align}
	S_{\mathrm{c.t.}} = \int d^4x \sqrt{-g}\,\left[ \alpha_1 R^2 + \alpha_2\left(R_{\mu\nu}R^{\mu\nu} 
	- \frac{1}{3}R^2\right)\right].
\end{align}
The term $R_{\mu\nu\rho\sigma}R^{\mu\nu\rho\sigma}$ is absorbed into the above two
terms up to the total derivative, and hence we ignore it in the following.
To the leading order in $h_{\mu\nu}$, the Ricci tensor is given as
\begin{align}
	R_{\mu\rho} \simeq \frac{1}{\Mp}\left[ \partial_\mu \partial_\rho {h^\alpha}_\alpha
	+ \partial^2 h_{\mu\rho} - \partial^\nu \partial_\rho h_{\mu\nu}
	- \partial^\nu \partial_\mu h_{\rho \nu}
	\right].
\end{align}
The Feynman rules are read off as
\begin{align}
	\begin{tikzpicture}[baseline=(a)]
	\begin{feynman} [inline]
		\vertex [label=270:\(\mu\nu\)] (a);
		\vertex [label=\(\alpha_1\), right = of a] (b);
		\vertex [label=270:\(\rho\sigma\), right = of b] (c);
		\diagram*{
		(a) -- [photon] (b) -- [photon, insertion={0}] (c)
		};
	\end{feynman}
	\end{tikzpicture}
	= \frac{8i \alpha_1}{\Mp^2} q^4 
	P_{\mu\nu} P_{\rho\sigma},
	\label{eq:grav_quad_alpha1}
\end{align}
for the $\alpha_1$-term, and
\begin{align}
	\begin{tikzpicture}[baseline=(a)]
	\begin{feynman}[inline]
		\vertex [label=270:\(\mu\nu\)] (a);
		\vertex [label=\(\alpha_2\), right = of a] (b);
		\vertex [label=270:\(\rho\sigma\), right = of b] (c);
		\diagram*{
		(a) -- [photon] (b) -- [photon, insertion={0}] (c)
		};
	\end{feynman}
	\end{tikzpicture}
	= \frac{i \alpha_2}{\Mp^2} q^4 
	\left[
	P_{\mu\rho}P_{\nu\sigma}
	+ P_{\mu\sigma} P_{\nu\rho}
	- \frac{2}{3} P_{\mu\nu} P_{\rho\sigma}
	\right],
\end{align}
for the $\alpha_2$-term,
where we define the projection operator as
\begin{align}
	P_{\mu\nu} = \eta_{\mu\nu} - \frac{q_\mu q_\nu}{q^2}.
\end{align}

\subsection{Tree-level amplitude}

We consider the 2-to-2 scattering amplitude $\phi_i \phi_i \rightarrow \phi_j \phi_j$ with $i\neq j$.
Only the $s$-channel process contributes in this case, which is given by
\begin{align}
	i A_\mathrm{tree}^{(ii\rightarrow jj)} &= 
	\begin{tikzpicture}[baseline=(c)]
	\begin{feynman}[inline]
		\vertex (c);
		\vertex [label=\(p_1{,} i\), above left = of c] (a);
		\vertex [label=270:\(p_2{,} i\), below left = of c] (b);
		\vertex [right = of c] (d);
		\vertex [label=\(p_3{,} j\), above right = of d] (e);
		\vertex [label=270:\(p_4{,} j\), below right = of d] (f);
		\diagram*{
		(a) -- [scalar] (c)
		(b) -- [scalar] (c)
		(c) -- [photon] (d)
		(d) -- [scalar] (e)
		(d) -- [scalar] (f)
		};
	\end{feynman}
	\end{tikzpicture}
	\nonumber \\
	&= \left(-\frac{i}{\Mp}\right) V_{\mu\nu}(p_1,p_2)
	\frac{i}{2\left(p_1+p_2\right)^2}P^{\mu\nu\rho\sigma}
	\left(-\frac{i}{\Mp}\right) V_{\rho\sigma}(p_3,p_4).
\end{align}
It is simplified as
\begin{align}
	A_\mathrm{tree}^{(ii\rightarrow jj)} 
	= \frac{1}{\Mp^2s}\left[\frac{\left(1+6\xi\right)^2}{6} s^2 - \left(\frac{s^2}{6}-tu\right)\right],
\end{align}
where the Mandelstam variables are defined as
\begin{align}
	s &\equiv \left(p_1+p_2\right)^2 = \left(p_3 + p_4\right)^2, \\
	t &\equiv \left(p_1-p_3\right)^2 = \left(p_2 - p_4\right)^2, \\
	u &\equiv \left(p_1-p_4\right)^2 = \left(p_2-p_3\right)^2.
\end{align}
It correctly reproduces the result in Ref.~\cite{Aydemir:2012nz}.

\subsection{Scalar one-loop amplitude}

Now we compute the one-loop correction to the 2-to-2 scattering amplitude.
We focus on the scalar loop here. It is given by
\begin{align}
	i A_\mathrm{1\mathchar`-loop}^{(ii\rightarrow jj)} &= 
	\begin{tikzpicture}[baseline=(c)]
	\begin{feynman}[inline]
		\vertex (c);
		\vertex [label=\(p_1\), above left = of c] (a);
		\vertex [label=270:\(p_2\), below left = of c] (b);
		\vertex [right = of c] (d);
		\vertex [right = of d] (e);
		\vertex [right = of e] (f);
		\vertex [label=\(p_3\), above right = of f] (g);
		\vertex [label=270:\(p_4\), below right = of f] (h);
		\diagram*{
		(a) -- [scalar] (c) -- [scalar] (b)
		(c) -- [photon] (d)
		(e) -- [scalar, half left, momentum=\(l-q\)] (d) -- [scalar, half left, momentum=\(l\)] (e)
		(e) -- [photon] (f)
		(g) -- [scalar] (f) -- [scalar] (h) 
		};
	\end{feynman}
	\end{tikzpicture}
	\nonumber \\
	=& \frac{N_s}{2}\int \frac{d^4 l}{\left(2\pi\right)^4}\left(-\frac{i}{\Mp}\right)^4\left(\frac{i}{2s}\right)^2
	\frac{i}{l^2 + i0} \frac{i}{\left(l-q\right)^2 + i0}
	\nonumber \\
	&\times \left[V_{\mu\nu}(p_1,p_2)P^{\mu\nu\rho\sigma}V_{\rho\sigma}(l,-l+q)\right]
	\left[V_{\alpha\beta}(l,-l+q)P^{\alpha\beta\gamma\delta}V_{\gamma\delta}(p_3,p_4)\right],
\end{align}
where $q = p_1 + p_2 = p_3 + p_4$, $s = q^2$ is the Mandelstam variable,
and the factor $1/2$ in front of the second line is the symmetry factor.
We first focus on the real (divergent) part.
By using the Feynman's trick,
\begin{align}
	\frac{1}{AB} = \int_0^1 dx \frac{1}{\left(A + (B-A)x\right)^2},
\end{align}
and moving to the Euclidean momentum space,
we obtain
\begin{align}
	\mathrm{Re}\left[A_{\mathrm{1\mathchar`-loop}}^{(ii\rightarrow jj)}\right] 
	=& \frac{N_s\mu^{4-d}}{2\Mp^4 s^2}
	\int_0^1 dx \int \frac{d^d l_E}{\left(2\pi\right)^d}\frac{1}{\left(l_E^2 + x(1-x)s\right)^2}
	\nonumber \\
	&\times \left[ 
	\left(1+4\xi\right)s l_E^2 - 4\left( \xi \left(q\cdot l_E \right)^2
	+ \left(p_1\cdot l_E\right)\left(p_2 \cdot l_E\right)\right)
	- s^2 \xi \left(1+6\xi\right)
	\right] \nonumber \\
	&\times \left[ \left(1+4\xi\right)s l_E^2 - 4\left( \xi \left(q \cdot l_E \right)^2
	+ \left(p_3\cdot l_E\right)\left(p_4 \cdot l_E\right)\right)
	- s^2 \xi \left(1+6\xi\right)
	\right],
\end{align}
where we have performed the dimensional regularization.
We can simplify the integrals as
\begin{align}
	\int \frac{d^d l_E}{\left(2\pi\right)^d}\,l_{E \mu} l_{E \nu} f(l_E^2) 
	&=
	\frac{\eta_{\mu\nu}}{d} \int \frac{d^d l_E}{\left(2\pi\right)^d}\,l_{E}^2 f(l_E^2), \\ 
	\int \frac{d^d l_E}{\left(2\pi\right)^d}\,l_{E \mu} l_{E \nu} l_{E \rho} l_{E\sigma} f\left(l_E^2\right) 
	&=
	\frac{\eta_{\mu\nu}\eta_{\rho\sigma} + \eta_{\mu\rho}\eta_{\nu\sigma}
	+\eta_{\mu\sigma}\eta_{\nu\rho}}{d\left(d+2\right)} 
	\int \frac{d^d l_E}{\left(2\pi\right)^d}\,l_{E}^4 f\left(l_E^2\right),
\end{align}
where $f$ is an arbitrary function of $l_E^2$.
After applying these formulas and some computation, 
we obtain the divergent part of the amplitude as
\begin{align}
	\left. A_\mathrm{1\mathchar`-loop}^{(ii\rightarrow jj)}\right\rvert_{\mathrm{div}}
	= \frac{N_s}{960\pi^2 \Mp^4}\frac{1}{\epsilon}\left[\frac{5}{6}\left(1+6\xi\right)^4 s^2 + 
	\left(\frac{s^2}{6} - tu\right) \right],
\end{align}
where $\epsilon$ is defined as $d = 4 - 2\epsilon$. It is renormalized by the $\alpha_1$- and 
$\alpha_2$-terms. The amplitude from the counter terms is
\begin{align}
	iA_{\mathrm{c.t.}}^{(ii\rightarrow jj)} &= 
	\begin{tikzpicture}[baseline=(c)]
	\begin{feynman} [inline]
		\vertex (c);
		\vertex [label=\(p_1\), above left = of c] (a);
		\vertex [label=270:\(p_2\),below left = of c] (b);
		\vertex [label=\(\alpha_1\), right = of c] (d);
		\vertex [right = of d] (e);
		\vertex [label=\(p_3\),above right = of e] (f);
		\vertex [label=270:\(p_4\),below right = of e] (g);
		\diagram*{
		(a) -- [scalar] (c) -- [scalar] (b)
		(c) -- [photon] (d) -- [photon, insertion={0}] (e)
		(f) -- [scalar] (e) -- [scalar] (g)
		};
	\end{feynman}
	\end{tikzpicture}
	+
	\begin{tikzpicture}[baseline=(c)]
	\begin{feynman} [inline]
		\vertex (c);
		\vertex [label=\(p_1\), above left = of c] (a);
		\vertex [label=270:\(p_2\),below left = of c] (b);
		\vertex [label=\(\alpha_2\), right = of c] (d);
		\vertex [right = of d] (e);
		\vertex [label=\(p_3\),above right = of e] (f);
		\vertex [label=270:\(p_4\),below right = of e] (g);
		\diagram*{
		(a) -- [scalar] (c) -- [scalar] (b)
		(c) -- [photon] (d) -- [photon, insertion={0}] (e)
		(f) -- [scalar] (e) -- [scalar] (g)
		};
	\end{feynman}
	\end{tikzpicture}
	\nonumber \\
	&= \frac{2i}{\Mp^4}
	\left[ \left(1+6\xi\right)^2 s^2 \alpha_1
	+ \left(\frac{s^2}{6} - tu\right) \alpha_2
	\right],
\end{align}
and hence the divergent parts of $\alpha_1$ and $\alpha_2$ are given by
\begin{align}
	\left. \alpha_1 \right\rvert_{\mathrm{div}} 
	&= - \frac{N_s \left(1+6\xi\right)^2}{2304\pi^2}\frac{1}{\epsilon}, 
	\label{eq:alpha1_div} \\
	\left. \alpha_2 \right\rvert_{\mathrm{div}} 
	&= - \frac{N_s}{1920\pi^2}\frac{1}{\epsilon}.
	\label{eq:alpha2_div}	
\end{align}
They coincide with Ref.~\cite{tHooft:1974toh} for $\xi = 0$.
Noting that the amplitude is associated with the factor 
$\mu^{4-d} \simeq 1 - \epsilon \ln \mu^{-2}$, we obtain the real part of the one-loop amplitude 
after the renormalization as
\begin{align}
	\mathrm{Re}\left[A_\mathrm{1\mathchar`- loop}^{(ii\rightarrow jj)}\right]
	= -\frac{N_s}{960\pi^2\Mp^4}\left[\frac{5}{6}
	\left(1+6\xi\right)^4s^2\ln\left(\frac{s}{\mu_1^2}\right) 
	+ \left(\frac{s^2}{6}-tu\right)\ln\left(\frac{s}{\mu_2^2}\right)\right].
\end{align}
Here $\mu_1$ and $\mu_2$ are the energy scales at which $\alpha_1$ and $\alpha_2$ 
are taken to vanish,
respectively.
They correspond to the parameter choices of our theory.
Next we consider the imaginary part. By the cutting rule, it is given by
\begin{align}
	\mathrm{Im}\left[A_\mathrm{1\mathchar`- loop}^{(ii\rightarrow jj)}\right]
	= \frac{N_s}{16 \Mp^4 s^2}
	&\int \frac{d^3 l_1}{\left(2\pi\right)^3 2l_1^0} \frac{d^3 l_2}{\left(2\pi\right)^3 2l_2^0}
	\left(2\pi\right)^4 \delta^{(4)}\left(l_1 + l_2 - q\right) \nonumber \\
	&\times
	\left[V_{\mu\nu}(p_1,p_2)P^{\mu\nu\rho\sigma}V_{\rho\sigma}(l_1,l_2)\right]
	\left[V_{\alpha\beta}(l_1,l_2)P^{\alpha\beta\gamma\delta}V_{\gamma\delta}(p_3,p_4)\right],
\end{align}
where the momenta $l_1$ and $l_2$ are now on-shell, i.e. $l_1^2 = l_2^2 = 0$.
The two-body phase space integral is reduced in the standard way as
\begin{align}
	\int \frac{d^3 l_1}{\left(2\pi\right)^3 2l_1^0} \frac{d^3 l_2}{\left(2\pi\right)^3 2l_2^0}
	\left(2\pi\right)^4 \delta^{(4)}\left(l_1 + l_2 - q\right)
	= \frac{1}{32\pi^2}\int d\Omega_l,
\end{align}
where $\Omega_l$ is the solid angle between $\vec{l}_1$ and $\vec{p}_1$.
After performing the integral, we obtain
\begin{align}
	\mathrm{Im}\left[A_\mathrm{1\mathchar`- loop}^{(ii\rightarrow jj)}\right]
	&=
	\frac{N_s}{960\pi \Mp^4}
	\left[\frac{5}{6}\left(1+6\xi\right)^4s^2
	+ \left(\frac{s^2}{6}-tu\right)\right].
\end{align}
By combining the real and imaginary parts, the one-loop amplitude is given by
\begin{align}
	A_\mathrm{1\mathchar`- loop}^{(ii\rightarrow jj)}
	= -\frac{N_s}{960\pi^2\Mp^4}\left[\frac{5}{6}
	\left(1+6\xi\right)^4s^2\left(\ln\left(\frac{s}{\mu_1^2}\right) - i\pi \right)
	+ \left(\frac{s^2}{6}-tu\right)
	\left(\ln\left(\frac{s}{\mu_2^2}\right) - i \pi \right)\right].	
\end{align}
It again agrees with Ref.~\cite{Aydemir:2012nz}.

\subsection{Renormalization group equation}
The scalar contributions to the RG equations for $\alpha_1$ and $\alpha_2$
are readily derived from Eqs.~\eqref{eq:alpha1_div} and~\eqref{eq:alpha2_div}.
The operators $R^2$ and $R_{\mu\nu}R^{\mu\nu}$ have the mass dimension of 4 in the $d$-dimensional 
spacetime, and hence the bare couplings associated with these operators 
have the mass dimension of $d-4 = -2\epsilon$.
Since the bare couplings do not depend on $\mu$,
we obtain
\begin{align}
	\beta_{\alpha_1} &= 
	-\frac{N_s}{1152\pi^2}\left(1+6\xi\right)^2, \\
	\beta_{\alpha_2} &=
	-\frac{N_s}{960\pi^2},
\end{align}
at the one-loop level.
They coincide with, e.g., (the scalar parts of) Ref.~\cite{Salvio:2014soa},
noting that their couplings $f_0^2$ and $f_2^2$ are related to $\alpha_1$ and $\alpha_2$
as $f_0^{-2} = 6\alpha_1$ and $f_2^{-2} = -\alpha_2$, respectively.
The RG equations of $\alpha_1$ and $\alpha_2$ are
also discussed, e.g. in Refs.~\cite{Fradkin:1981iu,Salvio:2014soa,Salvio:2015kka,
Codello:2015mba,Calmet:2016fsr,Markkanen:2018bfx,Ghilencea:2018rqg,
Elizalde:1993ee,Elizalde:1993ew,Myrzakulov:2016tsz} 
(although the sign of $\beta_{\alpha_1}$
is opposite in some references).

\section{Higgs scalaron inflationary model}
\label{app:HiggsR2}

In this appendix, we summarize basic properties of the Higgs scalaron inflationary model.

\subsection{Einstein frame action and unitarity}
\label{subsec:Einstein_HiggsR2}

We consider the following action:
\begin{align}
	S = \int d^4 x \sqrt{-g_J} 
	\left[
	\frac{\Mp^2}{2}R_J + \alpha_1 R_J^2
	+ \frac{\xi}{2} R_J \tilde{\phi}_i^2
	+ \frac{1}{2}g_J^{\mu\nu}\partial_\mu \tilde{\phi}_i \partial_\nu \tilde{\phi}_i
	- V\left(\tilde{\phi}_i\right)
	\right],
	\label{eq:action_HiggsR2}
\end{align}
where $i = 1, 2, ..., N_s$ and we add the subscript $J$ for the quantities in the Jordan frame.
With an auxiliary field $\tilde{\chi}$, the above action is equivalent to
\begin{align}
	S = \int d^4 x \sqrt{-{g}_J} 
	\left[
	\frac{\Mp^2}{2}{R}_J\left(1 + \frac{\xi \tilde{\phi}_i^2 + 4\alpha_1 \tilde{\chi}}{\Mp^2}\right)
	- \alpha_1 \tilde{\chi}^2
	+ \frac{1}{2}{g}_J^{\mu\nu}\partial_\mu \tilde{\phi}_i \partial_\nu \tilde{\phi}_i
	- V\left(\tilde{\phi}_i \right)
	\right].
	\label{eq:action_HiggsR2_aux}
\end{align}
One can indeed see that it returns to Eq.~\eqref{eq:action_HiggsR2} after integrating out $\tilde{\chi}$.
Now we perform the Weyl transformation,
\begin{align}
	{g}_{J\mu\nu} = \Omega^{-2} g_{E\mu\nu},
	~~~
	\Omega^2 = 1 + \frac{\xi \tilde{\phi}_i^2 + 4\alpha_1 \tilde{\chi}}{\Mp^2},
	\label{eq:weyl}
\end{align}
where the subscript $E$ indicates the Einstein frame.
The Ricci scalar is transformed as
\begin{align}
	R_J = \Omega^2 \left[R_E + \frac{3}{2}g^{\mu\nu}_E\partial_\mu \ln \Omega^2 \partial_\nu \ln \Omega^2
	- 3 \DAlambert_E \ln \Omega^2\right].
\end{align}
We redefine the fields as
\begin{align}
	s &\equiv \sqrt{\frac{3}{2}}\,\Mp \ln \Omega^2,
	~~~
	\phi_i \equiv \frac{\tilde{\phi}_i}{\Omega}.
	\label{eq:field_redef}
\end{align}
The action is given by
\begin{align}
	S = \int d^4 x \sqrt{-g_E}
	&\left[
	\frac{\Mp^2}{2} R_E + \frac{1}{2} \left(\partial s\right)^2
	+ \frac{1}{2}\left(\partial \phi_i + \frac{1}{\sqrt{6}}\frac{\phi_i}{\Mp}\partial s\right)^2
	\right. \nonumber \\
	&\left.
	-\frac{\Mp^4}{16 \alpha_1} 
	\left(1 - \exp\left(-\sqrt{\frac{2}{3}}\frac{s}{\Mp}\right) - \frac{\xi \phi_i^2}{\Mp^2}\right)^2
	- \frac{V\left( \Omega \phi_i \right)}{\Omega^4}
	\right].
	\label{eq:actionE_HiggsR2}
\end{align}
In particular, for the quartic potential $V(\tilde{\phi}_i) = \lambda \tilde{\phi}_i^4/4$, 
it is given by
\begin{align}
	S = \int d^4 x \sqrt{-g_E}
	&\left[
	\frac{\Mp^2}{2} R_E + \frac{1}{2} \left(\partial s\right)^2
	+ \frac{1}{2}\left(\partial \phi_i + \frac{1}{\sqrt{6}}\frac{\phi_i}{\Mp}\partial s\right)^2
	\right. \nonumber \\
	&\left.
	-\frac{\Mp^4}{16 \alpha_1} 
	\left(1 - \exp\left(-\sqrt{\frac{2}{3}}\frac{s}{\Mp}\right) - \frac{\xi \phi_i^2}{\Mp^2}\right)^2
	- \frac{\lambda}{4} \phi_i^4
	\right].
	\label{eq:actionE_HiggsR2_phi4}
\end{align}
This theory is unitary up to the Planck scale
since all the higher dimensional operators are suppressed by $\Mp$.
In particular, neither $\xi$ nor $\alpha_1$ shows up in the cut-off scale 
of the higher dimensional operators.
Still the perturbativity of the couplings requires
\begin{align}
	\frac{\xi^2}{4\alpha_1} \lesssim 4\pi.
\end{align}
These features are first pointed out in Ref.~\cite{Ema:2017rqn} 
and further studied in Ref.~\cite{Gorbunov:2018llf}.
They can be seen in the Jordan frame~\eqref{eq:action_HiggsR2_aux} as follows.
In order to discuss the cut-off scale and the size of the couplings,
we first have to make the fields canonically normalized.
The scalar fields $\tilde{\phi}_i$ are already canonical around $\tilde{\phi}_i = 0$,\footnote{
	If we instead expand $\tilde{\phi}_i$ around $\tilde{\phi}_i \neq 0$,
	the fields $\tilde{\phi}_i$ mix with the scalar part of the metric.
	In this case, we have to make them canonically normalized as well,
	as in Ref.~\cite{Bezrukov:2010jz}.
} but $\tilde{\chi}$ is not canonical as it does not have the standard kinetic term
in the Jordan frame. We have to redefine it as
\begin{align}
	\chi \equiv \frac{4\alpha_1\tilde{\chi} + \xi \tilde{\phi}_i^2}{\Mp},
\end{align}
because the kinetic term of $\tilde{\chi}$ is supplied from 
the non-minimal coupling to the Ricci scalar 
(that induces the kinetic mixing between $\tilde{\chi}$ and the scalar part of the metric).\footnote{
	The field $\chi$ is canonical only up to an $\mathcal{O}(1)$ factor, but
	is enough to estimate the cut-off scale and the perturbativity condition.
} The couplings $\xi$ and $\alpha_1$ now enter the action only in the form
\begin{align}
	\mathcal{L}_\chi = -\frac{\Mp^2}{16\alpha_1}\left(\chi - \frac{\xi \tilde{\phi}_i^2}{\Mp}\right)^2,
\end{align}
and hence they do not affect the cut-off scale.
It is also seen that the perturbativity requires $\xi^2/4\alpha_1 \lesssim 4\pi$.
The field redefinitions~\eqref{eq:weyl} and~\eqref{eq:field_redef},
and the resultant action~\eqref{eq:actionE_HiggsR2},
are a refined version of this argument,
since the Weyl transformation is essentially solving the kinetic mixing
between $\tilde{\chi}$ and the scalar part of the metric.

\subsection{Inflationary predictions}
\label{subsec:inflation}
Here we briefly summarize the inflationary prediction of
the Higgs scalaron model.
We may take the unitary gauge for the Higgs,
assuming that the Higgs has a large field value during inflation.
Then the action is given by Eq.~\eqref{eq:actionE_HiggsR2_phi4}
with $N_s = 1$. It contains two scalar fields,
but effectively reduces to a single field model
for $\xi, \alpha_1 \gg 1$,
as the other mode is heavy in this limit.\footnote{
	We assume that $\lambda$, $\xi$ and $\alpha_1$ are all positive.
}
After integrating out the heavy mode, the inflaton potential is given by
\begin{align}
	U(\phi) = \frac{\Mp^4}{4}\frac{1}{\xi^2/\lambda + 4\alpha_1}
	\left[1 - \exp\left(-\sqrt{\frac{2}{3}}\frac{\phi}{\Mp}\right)\right]^2,
\end{align}
where $\phi$ is the canonically normalized inflaton
which is an admixture of the scalaron and the Higgs.
The CMB normalization requires~\cite{Akrami:2018odb}
\begin{align}
	\frac{\xi^2}{\lambda} + 4\alpha_1 \simeq 2\times 10^9.
\end{align}
The spectral index $n_s$ and tensor-to-scalar ratio $r$ are given as
\begin{align}
	n_s \simeq 1 - \frac{2}{N_e},~~~
	r \simeq \frac{12}{N_e^2},
\end{align}
where $N_e$ is the number of e-folds after inflation.
An interesting feature of this model is that 
in principle we can determine $N_e$ precisely,
as the reheating dynamics is determined from the couplings to the standard model particles 
that are known, and
the cut-off scale is the Planck scale.
Indeed, Refs.~\cite{He:2018mgb,Bezrukov:2019ylq} 
have studied the reheating dynamics of this model in detail.
The former studies particle production at the first oscillation 
for a generic parameter space,
while the latter finds that the reheating temperature is as high as $10^{15}\,\mathrm{GeV}$
(or $N_e \simeq 59$) at least for a tuned parameter space.
A complete study on the reheating dynamics of this model is yet to be done.
Assuming that $N_e$ is between $50$ and $60$,
the spectral index is consistent with the current CMB observation~\cite{Akrami:2018odb},
and the tensor-to-scalar ratio is within the reach of the future observations~\cite{Abazajian:2016yjj}.
For more comprehensive analysis on the inflationary perturbation, 
see Refs.~\cite{Wang:2017fuy,Ema:2017rqn,Pi:2017gih,He:2018gyf,Gundhi:2018wyz,Enckell:2018uic}.

\subsection{Feynman rules around the vacuum}
\label{subsec:Feynman_HiggsR2}

Here we derive the Feynman rules of the Higgs scalaron model
around the vacuum $\phi_i = 0$ and $s = 0$
that are used in Sec.~\ref{subsec:pole}.
We expand the action~\eqref{eq:actionE_HiggsR2} around
$\phi_i = 0$ and $s = 0$ as 
\begin{align}
	S \simeq \int d^4 x \sqrt{-g}
	&\left[
	\frac{\Mp^2}{2}R + \frac{1}{2}\left(\left(\partial s\right)^2 + \left(\partial \phi_i \right)^2
	+ \frac{\partial \phi_i^2\,\partial s}{\sqrt{6}\,\Mp}\right)
	-\frac{m_s^2}{2} s^2
	+ \frac{\xi m_s}{\sqrt{8\alpha_1}} s \phi_i^2
	- \frac{\xi^2}{16\alpha_1} \phi_i^4
	\right],
\end{align}
where we have defined the scalaron mass as
\begin{align}
	m_s^2 = \frac{\Mp^2}{12\alpha_1},
\end{align}
and we have retained only the terms relevant for our computation.
As advertised, the scalaron mass squared is inversely proportional to $\alpha_1$.
The Feynman rules for the scalaron and $\phi_i$ are easily read off from this action.
We take the de~Donder gauge in the Einstein frame in this subsection.
The propagators for $\phi_i$ and the graviton are then given by Eqs.~\eqref{eq:scalar_prop}
and~\eqref{eq:graviton_prop}, respectively.
The propagator for the scalaron $s$ is
\begin{align}
	\begin{tikzpicture} [baseline=(a)]
	\begin{feynman}
		\vertex (a);
		\vertex [right= of a] (b);
		\diagram*{
		(a) -- (b)
		};
	\end{feynman}
	\end{tikzpicture}
	~= \frac{i}{p^2 - m_s^2 + i\epsilon}.
\end{align}
We denote the scalaron propagator by a solid line to distinguish it from $\phi_i$.
The scalar-scalar-scalaron vertex is given by
\begin{align}
	\begin{tikzpicture} [baseline=(c)]
	\begin{feynman} [inline]
		\vertex (c);
		\vertex [label=\(p_1{,} i\), above left = of c] (a);
		\vertex [label=270:\(p_2{,} j\), below left = of c] (b);
		\vertex [right = of c] (d);
		\diagram*{
		(a) -- [scalar] (c)
		(b) -- [scalar] (c)
		(c) -- (d)
		};
	\end{feynman}
	\end{tikzpicture}
	= i\left[\frac{\xi m_s}{\sqrt{2\alpha_1}} 
	+ \frac{1}{\sqrt{6}\,\Mp}\left(p_1 + p_2\right)^2\right]\delta_{ij},
\end{align}
where we have included the factor 2 from the exchange of the scalar fields,
and the momenta are both outgoing or both incoming.
The scalar four-point vertex is given by
\begin{align}
	\begin{tikzpicture} [baseline=(c)]
	\begin{feynman} [inline]
		\vertex (c);
		\vertex [label=\(p_1\), above left = of c] (a);
		\vertex [label=270:\(p_2\), below left = of c] (b);
		\vertex [label=\(p_3\), above right = of c] (d);
		\vertex [label=270:\(p_4\), below right = of c] (e);
		\diagram*{
		(a) -- [scalar] (c) -- [scalar] (b)
		(d) -- [scalar] (c) -- [scalar] (e)
		};
	\end{feynman}
	\end{tikzpicture}
	= \begin{cases}
	 \displaystyle -i \frac{3\xi^2}{2\alpha_1} 
	 & \mathrm{for}~~
	 \phi_i \phi_i \rightarrow \phi_i \phi_i, \vspace{2mm} \\
	 \displaystyle -i \frac{\xi^2}{2\alpha_1}
	 & \mathrm{for}~~ \phi_i \phi_i \rightarrow \phi_j \phi_j
	 ~~\mathrm{or}~~ \phi_i \phi_j \rightarrow \phi_i \phi_j
	 ~~\mathrm{with}~~ i\neq j,
	\end{cases}
\end{align}
where we have included the combinatory factors,
and ignored the contribution from the Higgs potential that is irrelevant for our discussion.
Finally the scalar-scalar-graviton vertex is given by Eq.~\eqref{eq:scalar_scalar_graviton}
with $\xi = 0$.
\bibliographystyle{utphys}
\bibliography{ref}

\end{document}